\documentclass[12pt, a4paper]{article}
\usepackage[a4paper,margin=2.54cm,left=2.54cm,right=2.54cm]{geometry}
\usepackage[english]{babel}

\usepackage{amsmath}
\usepackage{amssymb}
\usepackage{bm}
\usepackage{bbm}
\usepackage{graphicx}
\usepackage[colorlinks=true, allcolors=black]{hyperref}
\usepackage{authblk}

\usepackage{libertine}
\usepackage[libertine]{newtxmath}

\usepackage[caption=false]{subfig}

\usepackage{natbib}
\DeclareMathOperator*{\argmin}{arg\,min}

\usepackage{mathtools}

\DeclarePairedDelimiterX{\infdivx}[2]{(}{)}{
  #1\;\delimsize\|\;#2%
}
\newcommand{\infdiv}{D_{\text{KL}}\infdivx}

\title{Inferring ocean transport statistics with probabilistic neural networks}
\author{Martin T. Brolly}
\affil{School of Mathematics and Maxwell Institute for Mathematical Sciences, University of Edinburgh,
King’s Buildings, Edinburgh EH9 3FD, UK}
\begin{document}
\maketitle

\begin{abstract}

Using a probabilistic neural network and Lagrangian observations from the Global Drifter Program, we model the single particle transition probability density function (pdf) of ocean surface drifters. The transition pdf is represented by a Gaussian mixture whose parameters (weights, means and covariances) are continuous functions of latitude and longitude determined to maximise the likelihood of observed drifter trajectories. This provides a comprehensive description of drifter dynamics allowing for the simulation of drifter trajectories and the estimation of a wealth of dynamical statistics without the need to revisit the raw data. As examples, we compute global estimates of mean displacements over four days and lateral diffusivity. We use a probabilistic scoring rule to compare our model to commonly used transition matrix models. Our model  outperforms others globally and in three specific regions. A drifter release experiment simulated using our model shows the emergence of concentrated clusters in the subtropical gyres, in agreement with previous studies on the formation of garbage patches. An advantage of the neural network model is that it provides a continuous-in-space representation and avoids the need to discretise space, overcoming the challenges of dealing with nonuniform data. Our approach, which embraces data-driven probabilistic modelling, is applicable to many other problems in fluid dynamics and oceanography.
\end{abstract}

\section{Introduction}

The motion of turbulent fluids can be characterised usefully by dynamical statistics such as dispersion, energy spectra and velocity structure functions \citep[e.g.,][]{Batchelor1953, MoninYaglom}. In oceanography much effort has been directed towards inferring such statistics from observations~\citep[e.g.,][]{LaCasce2008, vanSebille}. In many cases, these inference tasks can be related to problems in conditional probability density estimation. For example, estimating single-particle dispersion is related to estimating the conditional density
\begin{align}\label{eq: dx_density}
    p\left(\bm{X}(t+\tau) - \bm{X}(t) \mid \bm{X}(t),\, t,\, \tau\right),
\end{align}
where $\bm{X}(t)$ is the position of a particle at time $t$, in that the dispersion is the variance of this distribution. Similarly, the velocity structure functions are moments of the conditional density
\begin{align}\label{eq: du_density}
    p\left(\bm{u}(\bm{x}_1,\, t) - \bm{u}(\bm{x}_2,\, t )\mid\bm{x}_1,\, \bm{x}_2,\, t\right),
\end{align}
where $\bm{u}(\bm{x},\, t)$ is the fluid velocity at position $\bm{x}$. By estimating full conditional densities like~\eqref{eq: dx_density} and~\eqref{eq: du_density}, it is possible to estimate simultaneously a number of related statistics. For instance,~\eqref{eq: dx_density} describes entirely the single-particle displacement statistics, while~\eqref{eq: du_density} encodes velocity structure functions of all orders, providing two-point Eulerian velocity statistics. It is no surprise, then, that estimating these conditional densities accurately is a nontrivial task.

In this work we consider a particular tool for conditional density estimation, the mixture density network (MDN)~\citep{Bishop}, and test its performance in learning fluid statistics from observations. MDNs are machine learning models, which combine artificial neural networks with probabilistic mixture models to represent conditional densities~\citep{Bishop2006}.
Their use has increased rapidly in recent years with applications in a variety of fields for a range of reduced order modelling and emulation tasks, including surrogate modelling of fluid flow~\citep{maulik2020probabilistic}, parameterisation of subgrid momentum forcing in ocean models~\citep{Guillaumin}, emulation of complex stochastic models in epidemiology~\citep{Davis2020} and multi-scale models of chemical reaction networks~\citep{Bortolussi}, and subgrid scale closures in large eddy simulations of turbulent combustion~\citep{Shin}.

We focus on learning the single-particle transition density~\eqref{eq: dx_density} in the ocean near-surface using Lagrangian trajectory data collected as part of the Global Drifter Program~\citep{Drifters}. A model of the transition density provides, at every point in the ocean, a probabilistic forecast for drifter displacements from that location.
We show that the MDN model outperforms existing stochastic models of drifter dynamics based on Ulam's method \citep{Ulam, Froyland2001}, as well as another simple benchmark model, and eliminates the difficulty of designing appropriate discretisations of space needed for such models.

From the transition density it is possible to derive estimates of a range of single-particle statistics. As examples, we provide maps of the mean displacement over four days as a function of initial position $\bm{X}_0$, as well as the lateral diffusivity.
The transition density produces highly non-Gaussian statistics in some regions. By calculating the Kullback--Leibler divergence between our full model and a simplified Gaussian model, we quantify and map non-Gaussianity in drifter displacements.

The MDN model also provides the basis for a discrete-time Markov process model of drifter dynamics, offering a continuous space alternative to Markov chain models which have been used in numerous studies~\citep{Maximenko, vanSebille2012, Miron, Miron2021}. We perform a global simulation of drifters for a period of ten years with initial positions given on a uniform grid, and reproduce the `garbage patches' in subtropical gyres seen in previous studies.

The article is structured as follows. In \S\ref{sec: cde} we discuss conditional density estimation and the estimation of conditional statistics.
In \S\ref{sec: mdn} we introduce MDNs. In \S\ref{sec: application} we describe the MDN model of the single-particle transition density from drifter observations. We compare its performance with alternative models, present derived single-particle statistics and simulate the clustering of drifters in subtropical gyres. In \S\ref{sec: discussion} we conclude and suggest further problems where MDNs may be a useful tool.

\section{Conditional modelling}\label{sec: cde}
While the aim of regression is to model $\mathbb{E}[\bm{Y}\mid \bm{X}]$, where $\bm{X}$ and $\bm{Y}$ are random variables, conditional modelling (or conditional density estimation, CDE) is the task of inferring the full conditional probability density $p(\bm{Y}\mid \bm{X})$~\footnote{We restrict attention to the case of continuous random variables.}. By modelling conditional densities, rather than just conditional means, we incorporate information about the variability of $\bm{Y}\mid \bm{X}$; more than this, conditional models can capture skewness, excess kurtosis and multimodality. This comprehensive description of conditional statistics is valuable in applications where single point-estimates are insufficient due to inherent variability, and where there is interest in non-Gaussian statistics, including those associated with rare events.
Conditional models can be used in two ways: (i) as stochastic surrogate models (or emulators), and (ii) as a tool for estimating conditional statistics.

Parametric conditional models (such as MDNs) assume that, for each possible value of $\bm{X}$, the distribution of $\bm{Y}\mid \bm{X}$ belongs to a certain family of parametric distributions, i.e.
\begin{align}\label{eq: parametric_CDE}
    p(\bm{Y}\mid \bm{X}) = \rho(\bm{Y}\,;\, \bm{\theta}(\bm{X})),
\end{align}
where $\rho({}\cdot{}\, ; \, \bm{\theta})$ is the probability density corresponding to a family of distributions parameterised by $\bm{\theta}$. In this case, not only must the form of $\rho$ be chosen, but the dependence on the conditioned variable must also be modelled by some representation of $\bm{\theta}(\bm{X})$.

\subsection{Estimating conditional statistics}
Given data $\{\bm{X}_i, \bm{Y}_i\}$, a standard approach to estimating conditional statistics $\mathbb{E}[\bm{f}(\bm{Y})\mid\bm{X}]$ is to first discretise (or `bin') in $\bm{X}$ and produce local estimates $\widehat{\mathbb{E}[\bm{f}(\bm{Y})]}(\tilde{\bm{X}})$ for each value of the discretised variable $\tilde{\bm{X}}$, typically by Monte Carlo estimation, such that
\begin{align}\label{discretised_estimates}
    \widehat{\mathbb{E}[\bm{f}(\bm{Y})]}(\tilde{\bm{X}}) := \frac{\sum_i \bm{f}(\bm{Y}_i)\,\mathbbm{1}_B(\bm{X}_i)}{\sum_i \mathbbm{1}_B(\bm{X}_i)},
\end{align}
where $\mathbbm{1}_B$ is the indicator function of $B$, the set of values of $\bm{X}$ whose discretised value is $\tilde{\bm{X}}$. For estimates~\eqref{discretised_estimates} to be useful, one must design a suitable discretisation of the domain of $\bm{X}$, which balances the need to choose a fine enough discretisation to resolve details in $\bm{X}$ with the need to take sufficiently large bins to have enough data for these estimates to have reasonably small variance. This can be especially challenging when data is sparse, or when the density of data is highly inhomogeneous.

\begin{figure}
    \centering
    \includegraphics[width=4.5in]{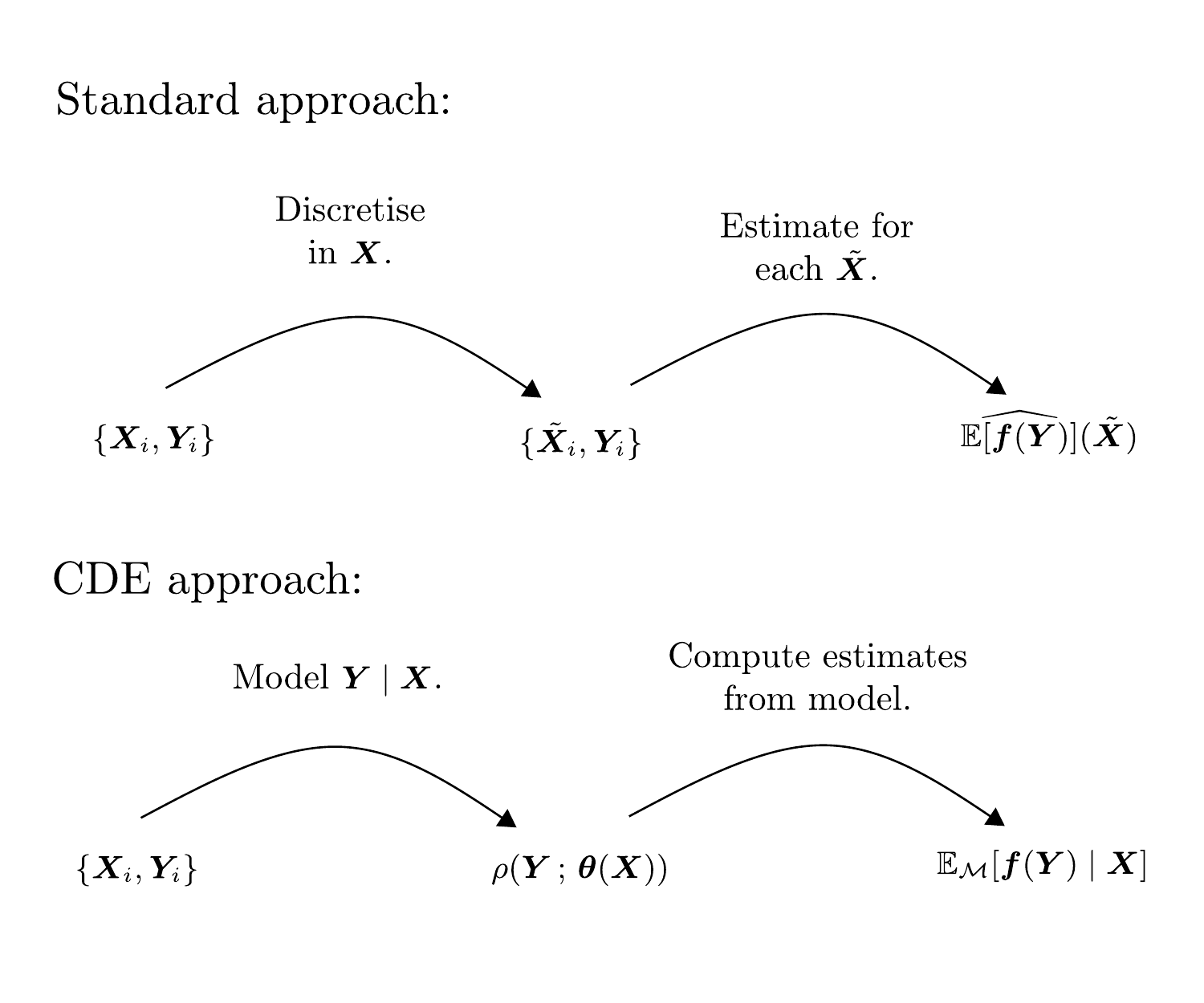}
    \caption{Estimating conditional statistics by a standard approach versus by first constructing a model for the conditional density.}
    \label{fig:cond_stats}
\end{figure}

Conditional modelling offers an alternative approach wherein one first constructs a model of the conditional density, as in \eqref{eq: parametric_CDE}, that is continuous in both $\bm{X}$ and $\bm{Y}$, then computes estimates 
\begin{align}
    \mathbb{E}_{\mathcal{M}}[\bm{f}(\bm{Y})\mid\bm{X}] := \int \bm{f}(\bm{Y})\ \rho(\bm{Y}\, ;\, \bm{\theta}(\bm{X}))\, \mathrm{d}\bm{Y}
\end{align}
for as many statistics as desired at any value of $\bm{X}$ in the domain, without the need to revisit the raw data. In some cases the expectations $\mathbb{E}_{\mathcal{M}}$ can be calculated using a closed-form expression. Where no such expression is known, the expectation can be computed by numerical integration or a Monte Carlo method. Since these calculations rely only on evaluating the modelled conditional density, or sampling from it, they are not limited by sparsity of data.
Also, for a given $\bm{X}^*$, estimates of the form~\eqref{discretised_estimates} are informed only by observations in the same bin as $\bm{X}^*$, whereas in a conditional model, all observations are used to fit $\rho(\bm{Y}\, ;\, \bm{\theta}(\bm{X}^*))$. The schematic in figure~\ref{fig:cond_stats} contrasts the standard approach and the conditional modelling approaches.

\section{Mixture density networks}\label{sec: mdn}
A mixture density network~\citep{Bishop, Bishop2006} is a conditional model where an artificial neural network is employed to represent the function $\bm{\theta}(\bm{X})$ in~\eqref{eq: parametric_CDE} and the parametric form $\rho({}\cdot{} ; \ \bm{\theta})$ corresponds to a mixture distribution. The density of a general mixture distribution is
\begin{align}\label{eq: mixture}
    \rho({}\cdot{};\ \bm{\theta}) = \sum_{i=1}^{N_c} \alpha_i\ \rho_i({}\cdot{};\ \bm{\theta}_i),
\end{align}
where $N_c$ is the number of components in the mixture, the $i^{\text{th}}$ component has density ${\rho_i({}\cdot{};\ \bm{\theta}_i)}$ with parameters $\bm{\theta}_i$, $\bm{\theta} = [(\alpha_1,\ \bm{\theta}_1),\ \cdots,\ (\alpha_{N_c},\ \bm{\theta}_{N_c})]$ and the $\alpha_i$ are component weights subject to the constraint
\begin{align}\label{eq: sum-to-one}
    \sum_{i=1}^{N_c}\alpha_i=1.
\end{align}
Commonly, the component densities $\rho_i$ are chosen from the same family and, in particular, Gaussian, but components can be chosen differently. In the Gaussian case, the $\bm{\theta}_i$ are conditional means and covariances.

The neural network representation of $\bm{\theta}({}\cdot{})$ is itself parametric with parameters $\bm{w}$; hence, MDNs model $p(\bm{Y}\mid\bm{X})$ with $\rho(\bm{Y}\, ;\, \bm{\theta}(\bm{X}\, ;\, \bm{w}))$.
The network can have any architecture, but that of a multilayer perceptron~\citep{Rumelhart} (also known as a fully connected multilayer feedforward neural network) with nonlinear activation functions is common --- in this case $\bm{w}$ consists of the weights and biases.

A natural loss function for conditional models, which quantifies how well they fit data, is the negative (conditional) log likelihood of observations $\mathcal{D}=\{\bm{X}_i, \bm{Y}_i\}$ under the model. In MDNs this is
\begin{align}\label{nll_loss}
    \mathcal{L}(\bm{w}\, ;\, \mathcal{D}) = \sum_i -\log\rho\left(\bm{Y}_i\, ;\, \bm{\theta}(\bm{X}_i\, ;\, \bm{w})\right).
\end{align}
Training an MDN then amounts to finding optimal values for the neural network's parameters
\begin{align}
    \bm{w}^* = \argmin_{\bm{w}}\,\mathcal{L}(\bm{w}\, ;\, \mathcal{D}).
\end{align}

Minimising the negative log likelihood is equivalent to maximising the log likelihood of training data, also referred to as the log score in probabilistic forecasting~\citep{Bernardo1979, GneitingRaftery, BroeckerSmith}. Maximum likelihood estimation in this context differs from the more familiar setting of fitting an unconditional model for $p(\bm{Y})$ given observed data $\{\bm{Y}_i\}$ --- here, there is generically only one observed value of $\bm{Y}\mid\bm{X}$ corresponding to each observed value of $\bm{X}$, and for most values of $\bm{X}$ there are no observations at all. It is clear, then, that, for each value of $\bm{X}$, we are certainly not in the large-data regime that would allow one to invoke asymptotic properties of maximum likelihood estimates. The quality of parametric conditional models~\eqref{eq: parametric_CDE} depends critically on how well $\bm{\theta}(\bm{X}\,;\,\bm{w}^*)$ represents how the distribution of $\bm{Y}\mid\bm{X}$ varies with $\bm{X}$. In particular, since MDNs employ a neural network to model $\bm{\theta}(\bm{X})$, and neural networks are highly flexible models, it is common for MDNs to exhibit poor generalisation unless regularisation techniques are used. In the following section we employ a widely used regularisation technique known as early stopping (see e.g.~\citet{Prechelt}), wherein a small proportion of training data (referred to as the test set) are not used to inform steps in the optimisation scheme, but are instead used to track the evolution of an estimate of the model's generalisation error (the value of the loss function evaluated on data outside the training set). The guiding heuristic is that it is typical for the generalisation error of neural networks to reach a minimum as training progresses before increasing due to overfitting --- early stopping is a strategy where one terminates model training when the generalisation error is believed to have reached this minimum. Details of our implementation are given in the following section.

\section{Application to single-particle statistics of the ocean near-surface}\label{sec: application}

In this section we present an MDN model of the single-particle transition density~\eqref{eq: dx_density} of ocean surface drifting buoys (drifters). The model's parameters are inferred from trajectory data collected as part of the Global Drifter Program~\citep{Drifters}. 

\subsection{Data}
\begin{figure}
    \centering
    \includegraphics[width=6in]{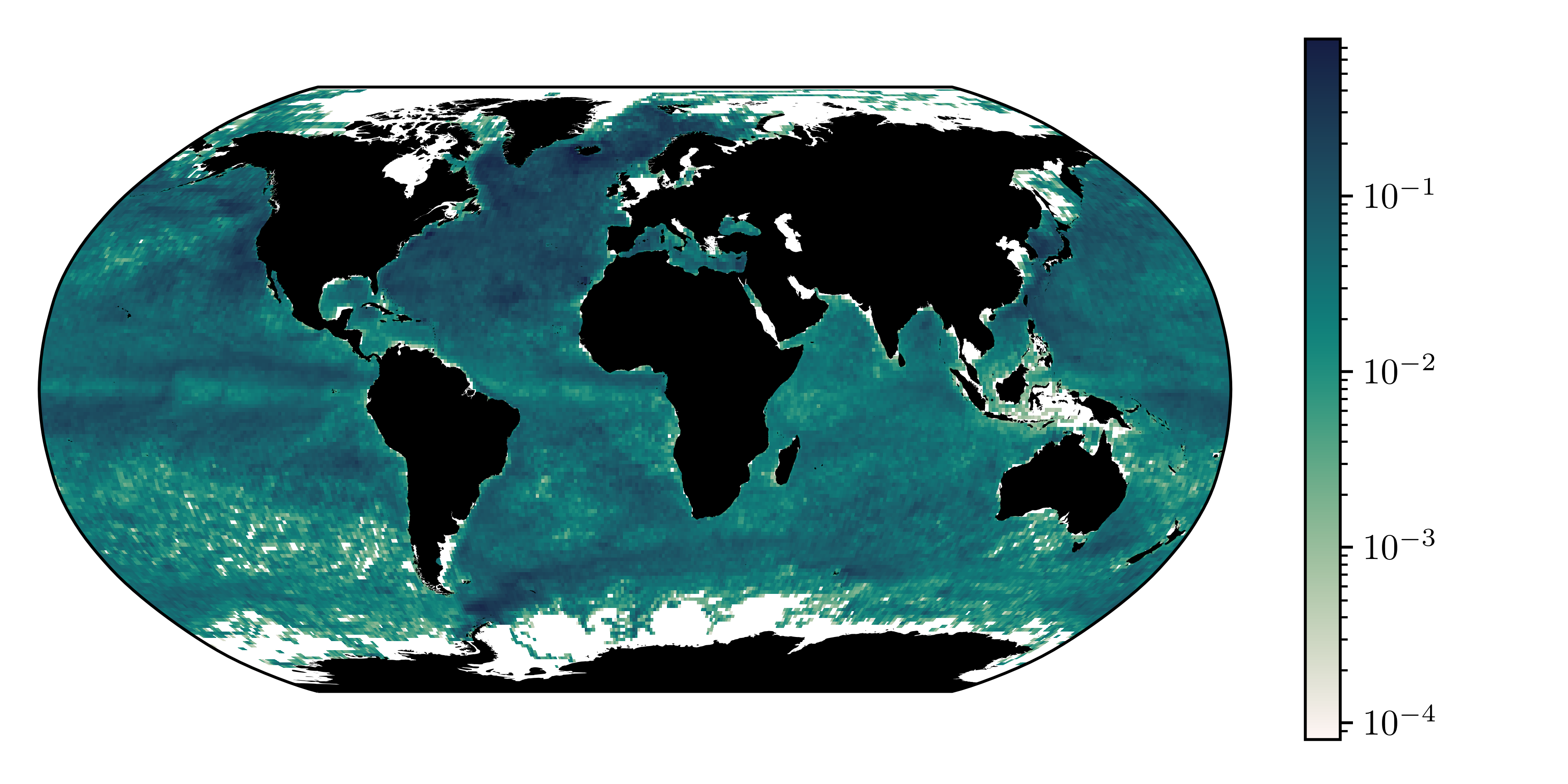}
    \caption{Count of drifter observations per squared kilometre.}
    \label{fig:drifters}
\end{figure}

We use the Global Drifter Program quality-controlled 6-hour interpolated dataset, which includes positions (latitude and longitude) and sea-surface temperatures. Drifter velocity estimates are also provided, though these are obtained by simple finite-differencing of position measurements. Position measurements are obtained from satellite fixes which are nonuniform in time and subject to error. The raw measurements are treated according to the procedure of~\citet{Hansen}, which involves the removal of suspected spurious values and interpolation to regular 6-hour intervals. The interpolation method, which is a form of kriging~\citep{HansenHerman}, assumes contamination by an uncorrelated zero-mean noise and makes assumptions about the structure functions of the discretised position process. We leave as a caveat to our results that this preprocessing of the data could be questioned and proceed taking the interpolated data as our ground-truth. Only position observations are used in our modelling. Figure~\ref{fig:drifters} shows how many observed displacements are recorded per squared kilometre in each $1^{\circ}\  \text{latitude} \ \times 1^{\circ} \ \text{longitude}$ square. These data were recorded between $1989$ and $2021$ and include a total of $23893$ drifter trajectories. We split the data in two parts, by selecting approximately half ($11946$) of the drifter trajectories at random to use for creating the model and set the remaining data aside for validation. The overall dataset contains over $18$ million observations of $6$-hour displacements.

\begin{figure}
    \centering
    \includegraphics{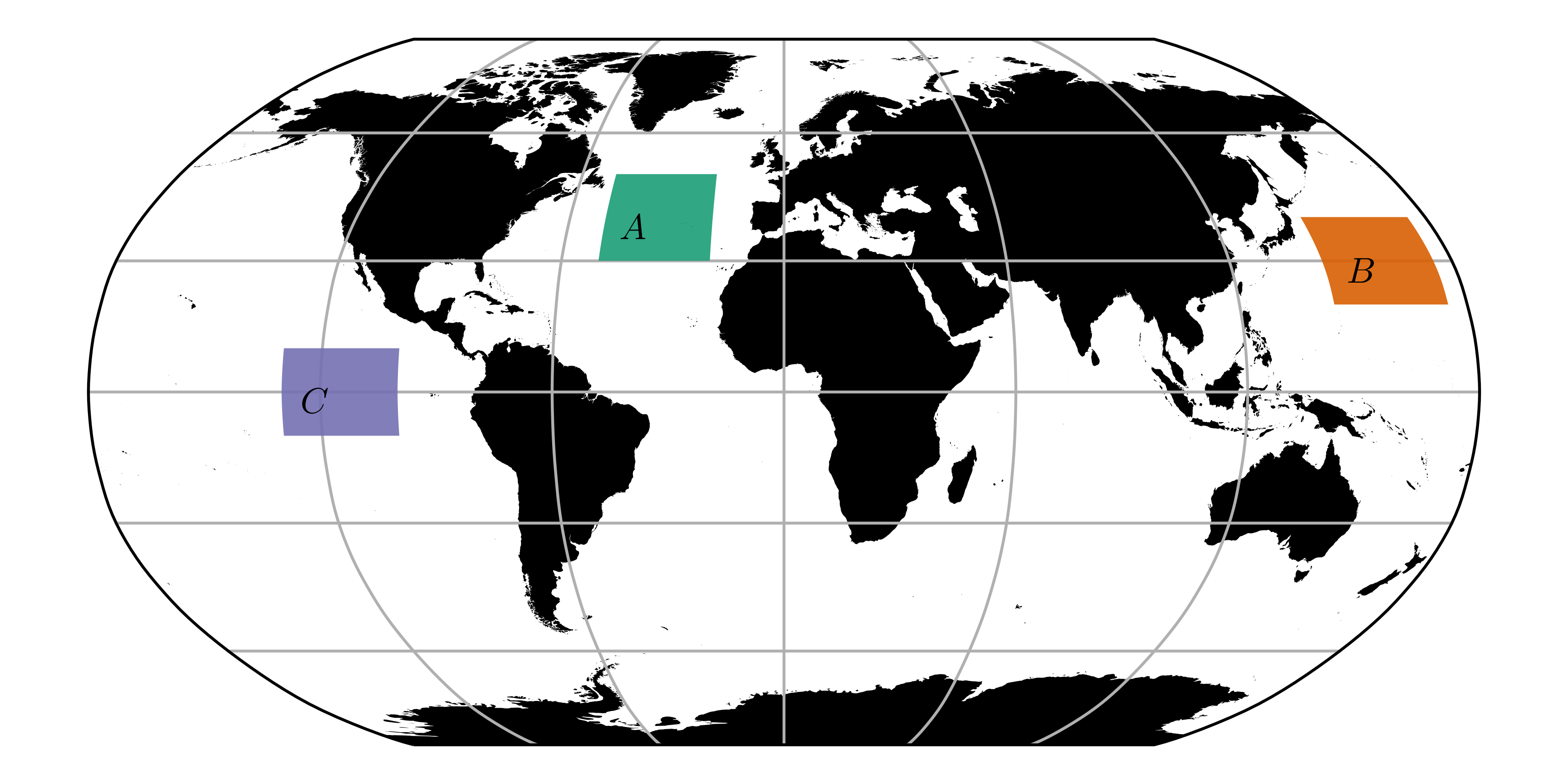}
    \caption{Regions considered for model comparison in section~\ref{model_comparison}.}
    \label{fig:regions}
\end{figure}

In section~\ref{model_comparison} we perform a model comparison. Skill scores are computed for the full training and validation datasets with global coverage, as well as for three restricted regions, labelled $A$, $B$, and $C$, shown in figure~\ref{fig:regions}, having extents $20$--$50^{\circ}$ W, $30$--$50^{\circ}$ N; $145$--$175^{\circ}$ E, $20$--$40^{\circ}$ N; and $110$--$130^{\circ}$ W, $10^{\circ}$ S--$10^{\circ}$ N.

\subsection{Model}
The transition density is not modelled in the most general form. Instead, we (i) consider, at first, a fixed value of the time-lag $\tau$, so that the transition density may be written
\begin{align}\label{eq: pXn}
    p(\bm{X}_{n+1}\mid \bm{X}_n),
\end{align}
where $\bm{X}_n = \bm{X}(t_0 + n\tau)$,
and (ii) assume the process $\bm{X}(t)$ is time-homogeneous, such that~\eqref{eq: dx_density} is independent of the initial time $t$, \eqref{eq: pXn} is independent of $n$ and 
\begin{align}\label{eq: pDX}
    p(\Delta\bm{X}\mid \bm{X}_0)
\end{align}
represents the same information as~\eqref{eq: pXn},
where $\Delta \bm{X}$ is the displacement of a drifter from its position at the previous timestep, denoted $\bm{X}_0$.
By assuming time-homogeneity we neglect the effects of seasonality and low-frequency variability in ocean dynamics.
If, additionally, an assumption of Markovianity is made, then \eqref{eq: pXn} is enough to construct a discrete-time Markov process model $(\bm{X}_n)$ for drifter position~\citep{Pav}. For a Markov assumption to be accurate, the discretisation timescale $\tau$ must be chosen appropriately. We choose a timescale of $4$ days on the basis that the Lagrangian velocity decorrelation time (or integral timescale) at the surface was previously estimated from drifters to be approximately $2$-$3$ days in all four ocean basins~\citep{Rupolo}.

\begin{figure}
    \centering
    \includegraphics[height=2in]{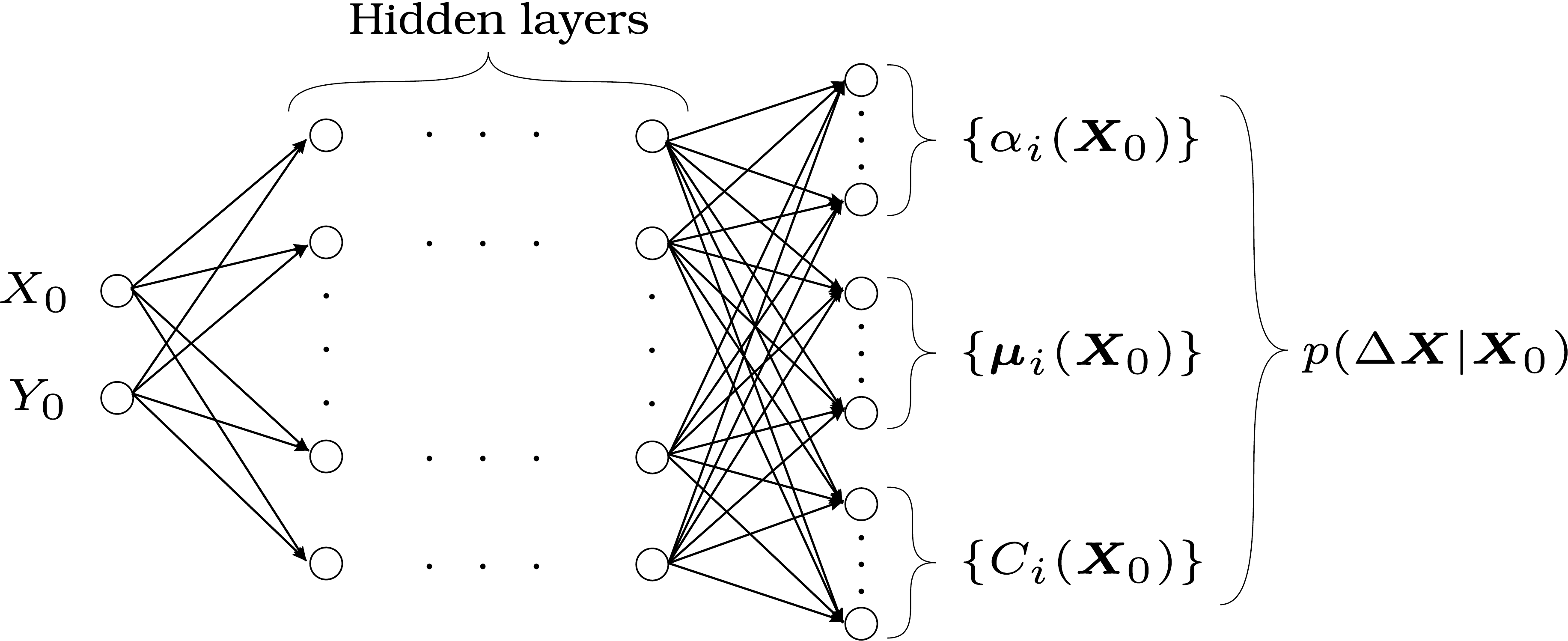}
    \caption{Schematic of the MDN model of the single-particle transition density of drifters.}
    \label{fig:mdn}
\end{figure}

We model~\eqref{eq: pDX} using an MDN ---  see the schematic in figure~\ref{fig:mdn}. The model takes as input $\bm{X}_0$, given in longitude--latitude coordinates, and its output is a Gaussian mixture distribution with $N_c=32$ mixture components modelling $\Delta \bm{X}\mid\bm{X}_0$, also in degrees of longitude and latitude from $\bm{X}_0$. The neural network part of the model thus encodes
\begin{align}
    \bm{\theta}(\cdot) = \big\{\alpha_i(\cdot),\  \bm{\mu}_i(\cdot),\  \bm{\mathsf{C}}_i(\cdot)\big\}_{i=1}^{N_c}
\end{align}
such that
\begin{equation}
    \begin{aligned}
        p(\Delta \bm{X}\mid \bm{X}_0) &= \sum_{i=1}^{N_c} \alpha_i(\bm{X}_0)\, \mathrm{det}\left(2\pi \bm{\mathsf{C}}_i\left(\bm{X}_0\right)\right) ^{-\frac{1}{2}}\,\\
            &\hspace{4em}\times\exp \left[-\frac{1}{2}\left(\Delta \bm{X}-\bm{\mu}_i\left(\bm{X}_0\right)\right)^\mathrm{T}\,\bm{\mathsf{C}}_i^{-1}\left(\bm{X}_0\right)\,\left(\Delta \bm{X}-\bm{\mu}_i\left(\bm{X}_0\right)\right)\right],
    \end{aligned}
\end{equation}
where $\bm{\mu}_i$ and $\bm{\mathsf{C}}_i$ are the mean vector and covariance matrix of mixture component $i$. The number of mixture components is a hyperparameter which could be optimised. We chose $N_c=32$ on the basis that $32$ component mixtures were found to be sufficiently expressive in trial experiments with MDNs.

The architecture chosen for the neural network is the standard multilayer perceptron, with six hidden (i.e. interior) layers. The first four hidden layers have 256 neurons and the remaining two have 512. The activation function $\tanh(x)$ is applied to each of the hidden layers. Thus, the activity of hidden layer $i$, is 
\begin{align}
    \bm{h}_i = \tanh\left(W_i\bm{h}_{i-1} + \bm{b}_i\right)
\end{align}
for $i>1$, and
\begin{align}
    \bm{h}_1 = \tanh\left(W_1\bm{X}_0 + \bm{b}_1\right).
\end{align}
Here, $\bm{W}_i\in \mathbb{R}^{d_{i}\times d_{i-1}}$ and $\bm{b}_i\in\mathbb{R}^{d_i}$ are the weight and bias parameters corresponding to the $i^{\text{th}}$ layer, having $d_i$ neurons. Note that $\bm{w}=\{\bm{W}_i, \bm{b}_i\}$.
The final layer has custom activation functions designed to enforce the natural constraints on the components of $\bm{\theta}$.
In particular, the softmax activation function $a_{\text{sm}}(\bm{x}) = \exp(\bm{x})\, /\, \sum_i\exp(x_i)$ is applied to the neural network outputs which correspond to the mixture component weights, $\bm{\alpha}$, to ensure that these are positive and satisfy the constraint~\eqref{eq: sum-to-one}.
Each covariance matrix $\bm{\mathsf{C}}_i$ is represented by the components of a lower triangular Cholesky factor --- positivity of the diagonal elements is enforced by taking an exponential. When $N_c=32$ we have $\text{dim}(\bm{\theta})=192$, and the total number of neural network parameters, i.e. weights and biases, is $\text{dim}(\bm{w}) = 690,880$. We train the model by minimising the negative log likelihood loss function~\eqref{nll_loss} using the Adam algorithm~\citep{Adam}. We note that the number of widths of hidden layers are further hyperparameters which we have chosen after experimentation with test problems. We do not attempt to find optimal values for these in this work.

As is common in machine learning, we standardise the data before training~\citep{LeCun2012}, that is we transform both the input data, $\{{\bm{X}_0}_i\}$, and output data, $\{{\Delta\bm{X}}_i\}$, separately, by subtracting the mean of the training data and dividing each component by its standard deviation in the training data, so that each component of the transformed data has zero mean and unit variance. While theoretical justifications for this practice are lacking or unsatisfactory, we found that it did improve noticeably the numerical stability of the optimisation procedure. In any case, the transformation that we apply is invertible, although care must be taken to correctly invert the rescaling of the transition density. For example, if we denote the standardised variables by $\widetilde{\bm{X}_0}$ and $\widetilde{\Delta\bm{X}}$, then the model approximates $p(\widetilde{\Delta\bm{X}}\mid\widetilde{\bm{X}_0})$, and we can recover the transition density with the correct units as
\begin{align}
    p(\Delta\bm{X}\mid\bm{X}_0) = \frac{p\left(\widetilde{\Delta\bm{X}}\mid\widetilde{\bm{X}_0}\right)} {\widehat{\text{std}}(\Delta X)\ \widehat{\text{std}}(\Delta Y)},
\end{align}
where $\widehat{\text{std}}({}\cdot{})$ denotes the sample standard deviation among the training data.

One aspect of neural networks that is particularly relevant to the problem at hand, is that they struggle to represent periodic functions~\citep{Liu}. Given that we operate in longitude--latitude coordinates, a model of the transition density ought to be periodic in longitude. However, since the neural network model receives the initial position $\bm{X}_0$ as simply a vector in $\mathbb{R}^2$, the concept of a spherical domain is not built in to the representation. Indeed, the MDN model produces discontinuities in $p(\Delta\bm{X}\mid\bm{X}_0)$ at the dateline due to model error on either side. To improve continuity at the dateline we employ a crude technique, wherein we replicate the data twice, once shifted by $360^{\circ}$ longitude west, and once shifted by $360^{\circ}$ east.

\begin{figure}
    \centering
    \includegraphics[width=4in]{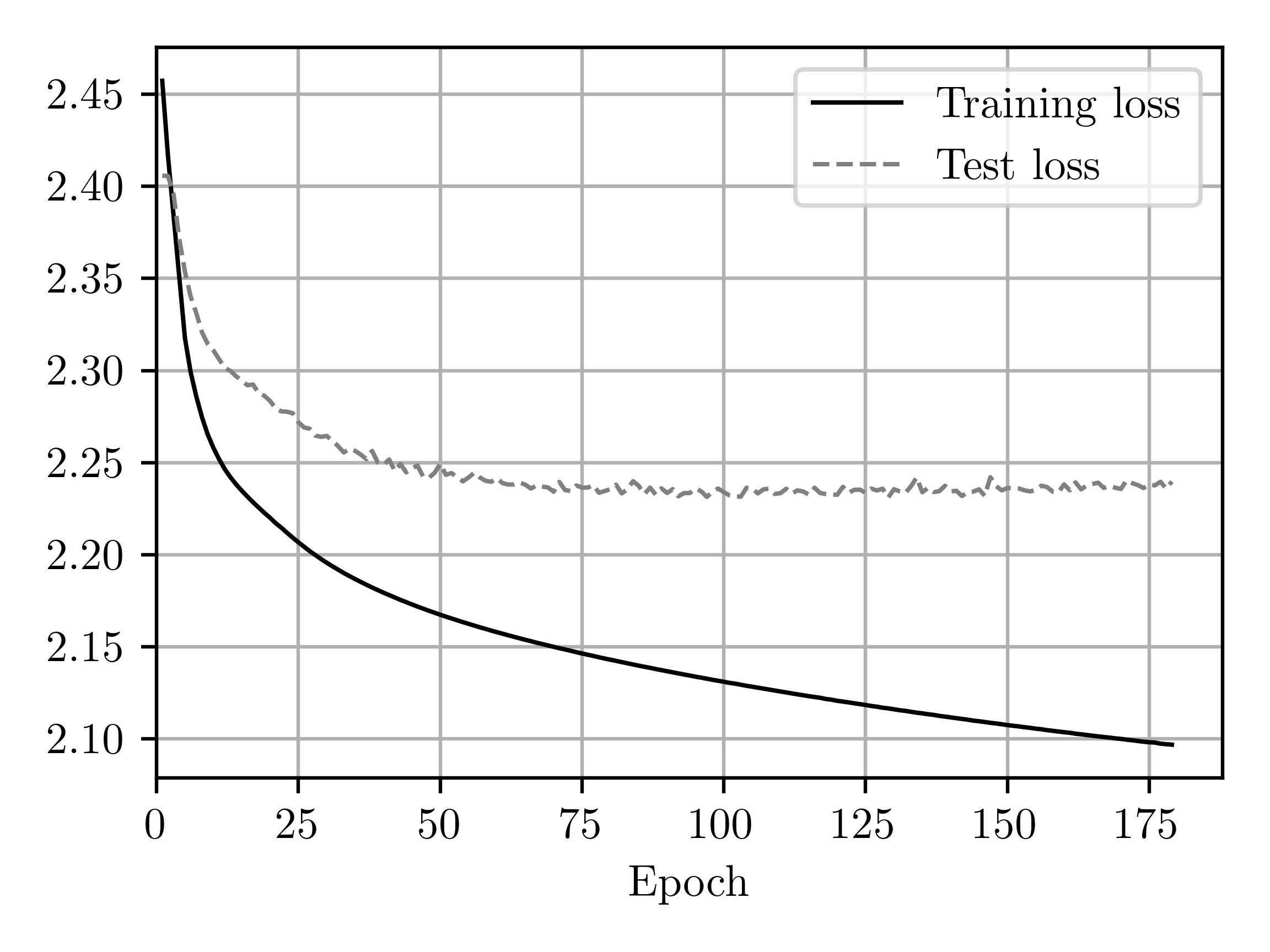}
    \caption{Evolution of the training and test loss in the MDN model during optimisation. The loss shown is the mean negative log likelihood per datapoint (i.e. a normalised form of \eqref{nll_loss}) in terms of the standardised variables $\widetilde{\bm{X}_0}$ and $\widetilde{\Delta\bm{X}}$.}
    \label{fig:loss}
\end{figure}

The model is implemented in Python using TensorFlow~\citep{tensorflow} and TensorFlow Probability~\citep{tensorflow-prob} and trained using four NVIDIA Tesla V100 16GB GPUs in parallel.
Of the $50\%$ of data used to construct the model, $90\%$, again chosen randomly, was used to inform iterations of the optimisation procedure, and $10\%$ was used for early stopping ---  we refer to these portions of the data as the training and test sets, respectively. Training took approximately $90$ minutes. The evolution during training of the loss function on training and test sets is shown in figure~\ref{fig:loss} as a function of epoch. An epoch is the number of iterations taken for all data to be used once in the Adam algorithm. The stopping criterion used for the optimisation, an example of early stopping, was that the test loss had not decreased since $50$ epochs previous.

\subsection{Model evaluation and comparison}\label{model_comparison}

Since the MDN model is probabilistic, its performance should be assessed using skill scores for probabilistic forecasts, as opposed to performance metrics commonly used for deterministic models, such as the mean squared error. As discussed above, minimising the negative log likelihood is equivalent to maximising the log score, since this is exactly the log likelihood. The log score has attractive properties, namely strict propriety~\citep{BroeckerSmith} and locality~\citep{Du2021}. Indeed it is the only smooth local strictly proper scoring rule for continuous variables up to affine transformation~\citep{Bernardo1979}. A scoring rule is strictly proper if its expectation (with respect to data) is maximised uniquely by the correct/perfect model (assuming it exists). A scoring rule is local if it is a function only of the value of the forecast probability distribution evaluated at the observed data, and does not depend for example on other features of the forecast distribution, such as its shape. For validation purposes we can compute the log score on our validation data set. However, while the value of the log score can be easily interpreted in the case of forecasts of discrete/categorical variables, its value in the case of continuous variables is not immediately meaningful, since it refers to probability density, which has dimensions inverse to the area of its support, meaning that the scale of the log score is problem-dependent. On the other hand, the log score can be more easily interpreted when used as a relative score between models --- in particular, the mean difference of log scores between models reflects the average additional probability the first model places on observed outcomes compared to the other model, measured in units of information, nats (or shannons when $\log_2$ is used in the definition of the score). The difference of log scores is invariant under smooth transformations of the forecast variable~\citep{Du2021}; this means, in particular, that differences in log scores are unaffected by a change of units. Thus, in order to evaluate the MDN model we compare it with alternative models.  We describe two alternative models, one used extensively in the literature, and one proposed here as a simple but reasonable alternative. We also compare with a simplified version of the MDN model, which features only one mixture component, i.e. for which $N_c=1$. The log score of all models is computed on both training and validation data to assess relative performance.

\subsubsection{Transition matrix model}\label{markov_chain}

Previous work~\citep{Maximenko, vanSebille2012, Miron, Miron2021}, modelled drifter dynamics with a discrete-time Markov chain using Ulam's method~\citep{Ulam, Froyland2001}. This requires to discretise space into bins $\{B_i\}$ and estimate the transition matrix
\begin{align}
    P_{ij} = \mathbb{P}(\bm{X}_{n+1} \in B_j \mid \bm{X}_n \in B_i),
\end{align}
which is the discrete analogue of the transition density~\eqref{eq: pXn}.
Indeed the primary difference between a Markov chain model and our Markov process model is that ours is continuous in space. The elements of the transition matrix are usually estimated by the standard approach sketched in figure~\ref{fig:cond_stats}, where we have $\bm{Y}=\bm{X}_{n+1}$ and $f(\bm{Y})=\mathbbm{1}_{B_j}(\bm{X}_{n+1})$ --- this corresponds to the maximum-likelihood estimate for each $P_{ij}$ and, hence, maximises the log score on the training dataset. Note that the transition matrix can be used to construct a corresponding transition density which is piecewise constant on gridcells in $\bm{X}_n$ and $\bm{X}_{n+1}$ via
\begin{align}\label{eq:mass_to_density}
    p(\Delta \bm{X}\mid\bm{X}_0) = \frac{P_{ij}}{A(B_j)},\quad \text{when}\ \bm{X}_0\in B_i,\ \bm{X}_0 + \Delta\bm{X} \in B_j,
\end{align}
where $A(B_j)$ is the area\footnote{For consistency with the transition density as given by the MDN model, these areas must be calculated in terms of the same variables, i.e. degrees longitude by degrees latitude.} of $B_j$. This is important for allowing comparison with models which are continuous in space.

An advantage of Markov chain models is that analysis of their long time behaviour is straightforward --- the left and right eigenvectors of the transition matrix can be studied to identify almost-invariant sets, as in~\citet{Miron}. This has been called the eigenvector method~\citep{Froyland2014}. The extension of this analysis to the continuous-space setting using our model, which we leave for future work, requires the calculation of eigenfunctions of the relevant Perron--Frobenius operator $\mathcal{P}$, which acts on probability density functions to evolve them forward in time, such that
\begin{align}
    p(\bm{X}_{n+1}) &= \mathcal{P}(p(\bm{X}_{n}))\\
    :&= \int_{\Omega}\, p(\bm{X}_n)\, p(\bm{X}_{n+1}\mid \bm{X}_n)\, \mathrm{d}\bm{X}_n.
\end{align}
Alternatively, it is worth noting that the MDN model can be used to construct a transition matrix, by numerical integration of the transition density, that is by computing numerically
\begin{align}
    P_{ij} = \int_{B_{i}}\int_{B_{j}} p(\bm{X}_{n+1}\mid \bm{X}_n)\, \mathrm{d}\bm{X}_{n+1}\, \mathrm{d}\bm{X}_n.
\end{align}

\begin{figure}
    \centering
    \includegraphics[width=6.2in]{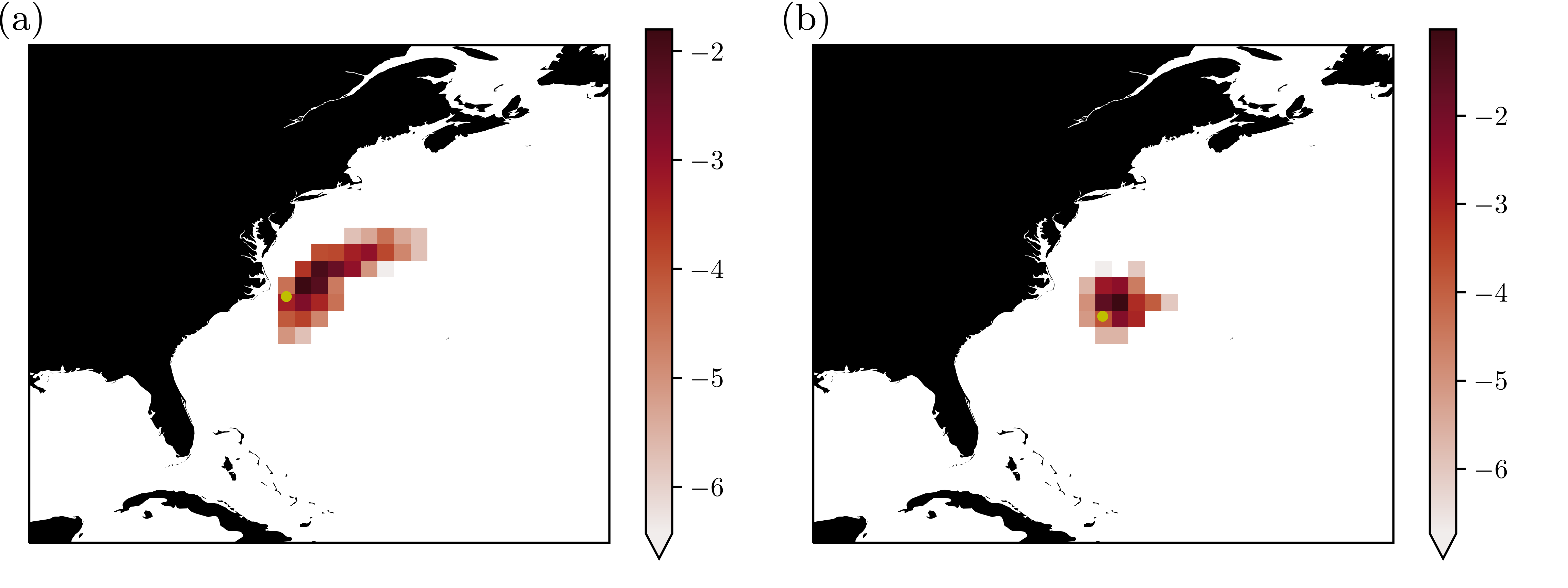}
    \caption{Maps of the log transition probability density function, $\log p(\Delta\bm{X}\mid\bm{X}_0)$, for initial positions, $\bm{X}_0$, (a) in the Gulf Stream ($34.85^{\circ}$~N,~$74.50^{\circ}$~W), and (b) adjacent to the Gulf Stream ($33.67^{\circ}$~N,~$72.55^{\circ}$~W), derived from the transition matrix model with $\tau=4$ days via~\eqref{eq:mass_to_density}. Yellow dots indicate $\bm{X}_0$.}
    \label{fig:DTMC_gulf_stream}
\end{figure}

In figure~\ref{fig:DTMC_gulf_stream} we show the log transition density $\log p(\Delta\bm{X}\mid\bm{X}_0)$ derived from the transition matrix model via~\eqref{eq:mass_to_density} for two different initial positions $\bm{X}_0$. The first is located within the core of the Gulf Stream at $34.85^{\circ}$~N,~$74.50^{\circ}$~W, and the second is just outside the Gulf stream at $33.67^{\circ}$~N,~$72.55^{\circ}$~W. Notice that in each case the support of the density is the set of grid cells to which transitions were observed in the training data. In other words, transitions to other grid cells have probability zero under the model. We return to this point in section~\ref{sec: scores}.

\subsubsection{Gaussian transitions with gridded parameters (GTGP)}
A simple model for the transition density~\eqref{eq: pDX} is that, given initial positions $\bm{X}_0$, transitions are conditionally Gaussian with conditional mean and covariance given by functions of $\bm{X}_0$ which are piecewise constant on grid cells, i.e.
\begin{align}
    \Delta\bm{X}\mid\bm{X}_0 \sim \mathcal{N}\left(\bm{\mu}\left(\bm{X}_0\right),\, \bm{\mathsf{C}}\left(\bm{X}_0\right)\right),
\end{align}
with $\bm{\mu}\left(\bm{X}_0\right)$ and $\bm{\mathsf{C}}\left(\bm{X}_0\right)$ piecewise constant in $\bm{X}_0$. The parameters $\bm{\mu}$ and $\bm{\mathsf{C}}$ are estimated by sorting the observations into bins and computing the sample mean and sample covariance for each bin. The sample mean is the maximum likelihood estimate of $\bm{\mu}$, while the sample covariance differs from the maximum likelihood estimate of $\bm{\mathsf{C}}$ only by a factor of $\frac{N - 1}{N}\approx 1$, where $N$ is the number of training data in the given bin. Hence, the parameter estimates used are very close to those which maximise the log score on training data.

\begin{figure}[htp]
    \centering
    \subfloat[Mean of 4-day zonal displacement (km).]{
        \includegraphics[clip, width=6in]{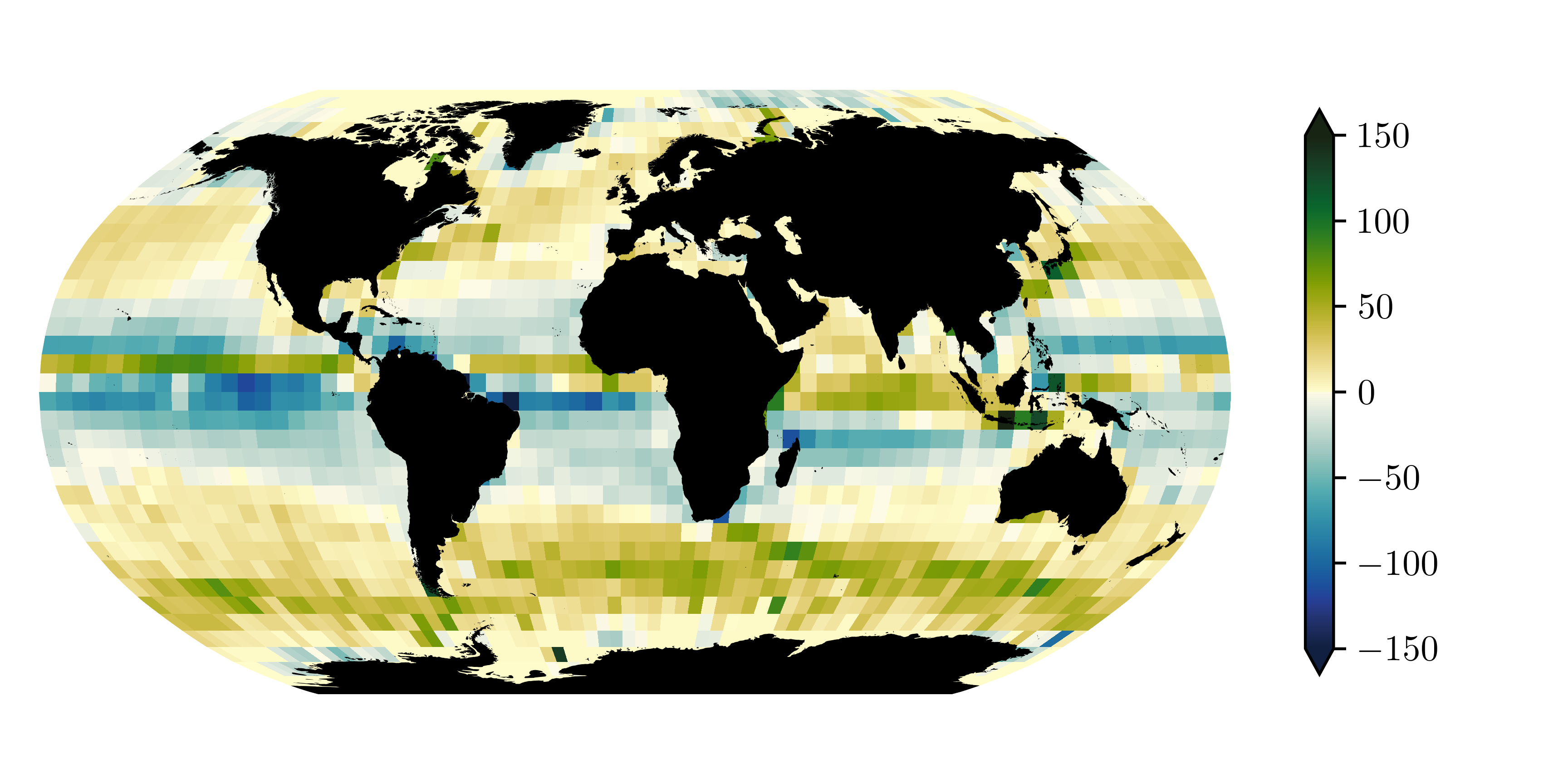}}
    \hspace{0pt}
    \subfloat[Mean of 4-day meridional displacement (km).]{
        \includegraphics[clip, width=6in]{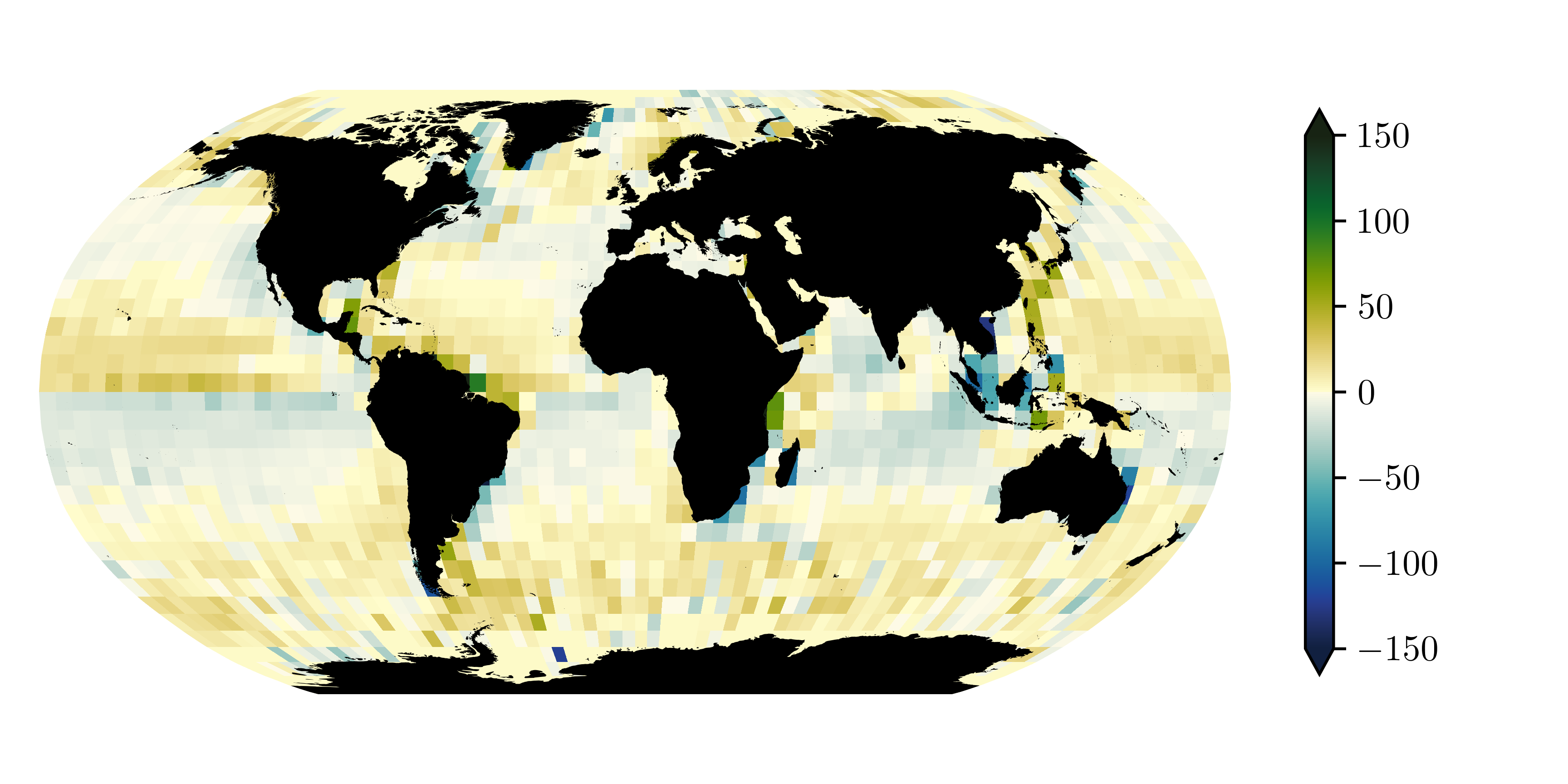}}
    \caption{Mean of displacements from the GTGP model, with $\tau =4\text{ days}$, as a function of initial position.}
    \label{fig:GTGP_mean}
\end{figure}

Figure~\ref{fig:GTGP_mean} shows the mean of displacements from a GTGP model with a regular $1^{\circ}\times1^{\circ}$ longitude--latitude grid and $\tau =4\text{ days}$, as a function of initial position.

\subsubsection{Model scores}\label{sec: scores}
We compute skill scores for the full training and validation datasets with global coverage, and for regions $A$, $B$, and $C$.  For both the transition matrix and GTGP models it is necessary to choose a spatial discretisation; herein we consider only square latitude--longitude grids, so that the only parameter to be chosen is the grid cell side length. This choice affects their performance. If a relatively high resolution discretisation is used, the models attain relatively high scores in training, but generalise poorly, as reflected in poor scores on validation data. In the case of the GTGP model, an issue arises when validation data falls in grid cells not visited by drifters in the training set, since sample means and covariances cannot be estimated in bins where data is absent. As a simple solution, we set the value of $\bm{\mu}$ and $\bm{\mathsf{C}}$ on unvisited grid cells equal to a global (or regional) estimate. A flaw of the transition matrix model, with the transition matrix estimated as discussed above, is that validation data can have zero probability under the model, and hence achieve a log score of minus infinity. This situation is avoided by taking sufficiently large grid cells, but this leads to exceptionally low scores.
On the other hand, a validation score can be computed with smaller grid cells if one is prepared to simply discard validation data which have zero probability under the model. This seems overly generous, as the transition matrix model will be scored increasingly highly as the grid cell size is reduced to zero and an increasing number of the validation data are neglected. As a compromise, we fix the grid cell size for the transition matrix model to $1^{\circ}\times1^{\circ}$, the resolution used in some previous studies~\citep{vanSebille2012} where the transition matrix model was used, and discard validation data with zero probability --- the proportion of validation data discarded was $7\%$ globally, and $2\%$, $9\%$ and $4\%$ in regions $A$, $B$ and $C$, respectively.
For the GTGP model the grid cell size was optimised to maximise validation scores using grid search cross validation: a common procedure which amounts to trying a range of values of a model hyperparameter (in this case the grid cell size) and choosing the value which optimises the validation score. The optimal grid cell size found ranged from $1.1^{\circ}$ in region A to $5^{\circ}$ globally. The scores are presented in table~\ref{tab:scores}. In all regions the MDN models outperform the alternatives, with the $32$-component model achieving slightly higher scores than the single-component model. Note that the scores reported happen to be negative --- this is not by convention, but instead reflects that log probability densities are often negative. A higher score is a better score.


\begin{table}
\centering
\begin{tabular}{ll|lllll}
    && TM  &   GTGP    &   MDN1    &   MDN32 \\ \hline
    Global: & Training & $-1.61$ & $-1.20$ & $-1.07$ & $-1.02$ \\
                & Validation & $-1.74$& $-1.30$ & $-1.14$ & $-1.11$ \\
    $A$: & Training & $-1.78$ & $-1.25$ & $-1.35$ & $-1.30$ \\
                & Validation & $-1.89$ & $-1.42$ & $-1.40$ & $-1.35$ \\
    $B$: & Training & $-1.95$ & $-1.90$ & $-1.91$ & $-1.85$ \\
                & Validation & $-2.16$ & $-2.01$ & $-2.01$ & $-1.96$ \\
    $C$: & Training & $-1.93$ & $-1.59$ & $-1.63$ & $-1.56$ \\
                & Validation & $-2.00$ & $-1.66$ & $-1.61$ & $-1.57$ \\
    
\end{tabular}
\caption{Training and validation scores for the transition matrix and GTGP models, as well as the single-component MDN and full $32$-component MDN models, in each of the regions considered (see the map in figure~\ref{fig:regions}). The scores are the mean log score (i.e. mean log likelihood) per datapoint calculated in terms of the variables $\bm{X}_0$ and $\Delta\bm{X}$ in their original degrees longitude/latitude units.}
\label{tab:scores}
\end{table}

\subsection{Results}
\begin{figure}
    \centering
    \includegraphics[width=6.2in]{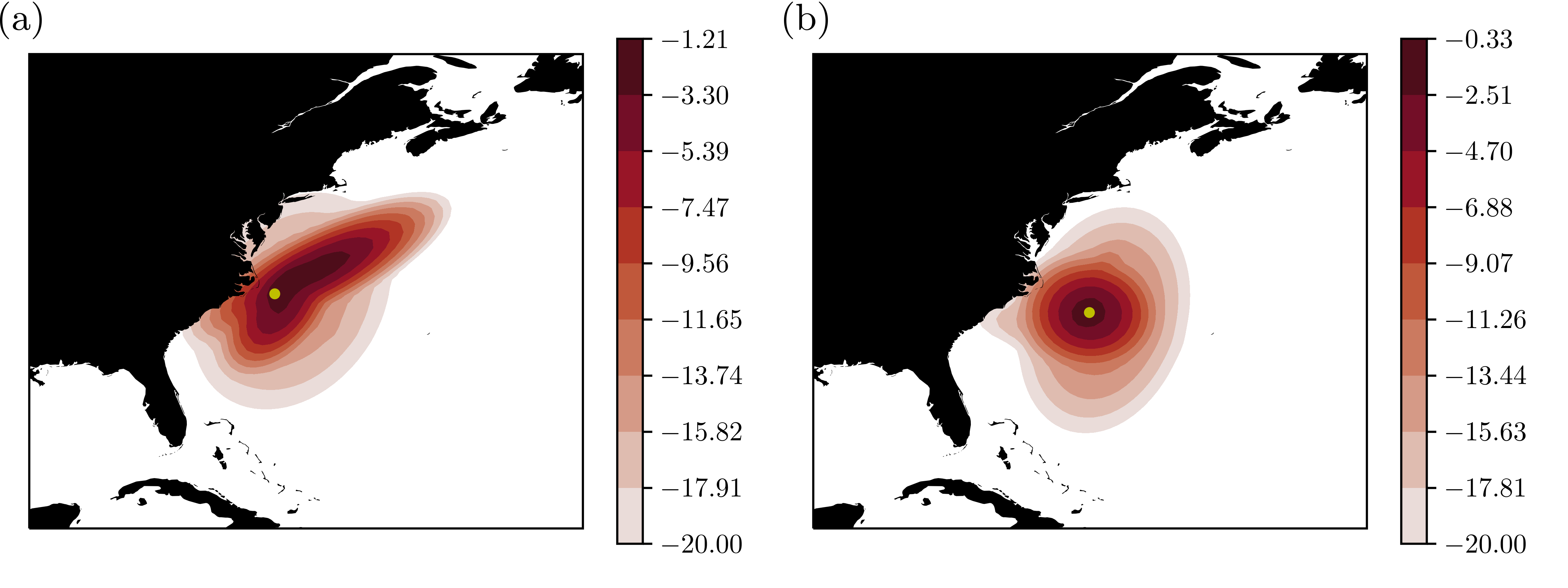}
    \caption{Maps of the log transition density, $\log p(\Delta\bm{X}\mid\bm{X}_0)$, for initial positions, $\bm{X}_0$, (a) in the Gulf Stream ($34.85^{\circ}$~N,~$74.50^{\circ}$~W), and (b) adjacent to the Gulf Stream ($33.67^{\circ}$~N,~$72.55^{\circ}$~W), derived from the MDN model with $\tau=4$ days. Yellow dots indicate $\bm{X}_0$.}
    \label{fig:gulf_stream}
\end{figure}

Once trained, the model can be used in at least two ways: (i) to derive estimates of single-particle displacement statistics, and (ii) to simulate drifter trajectories. However, we first examine the transition density directly. In figure~\ref{fig:gulf_stream} we show the log transition density $\log p(\Delta\bm{X}\mid\bm{X}_0)$ for two different initial positions $\bm{X}_0$. In the first case, where $\bm{X}_0$ is located within the core of the Gulf Stream at $34.85^{\circ}$~N,~$74.50^{\circ}$~W, the transition density is strongly non-Gaussian, with contours extending roughly to the south and northeast, showing the influence of the Gulf Stream on drifters. In the second case, where $\bm{X}_0$ is just outside the Gulf stream at $33.67^{\circ}$~N,~$72.55^{\circ}$~W, the transition density is closer to Gaussian.

\begin{figure}
    \centering
    \includegraphics[width=6in]{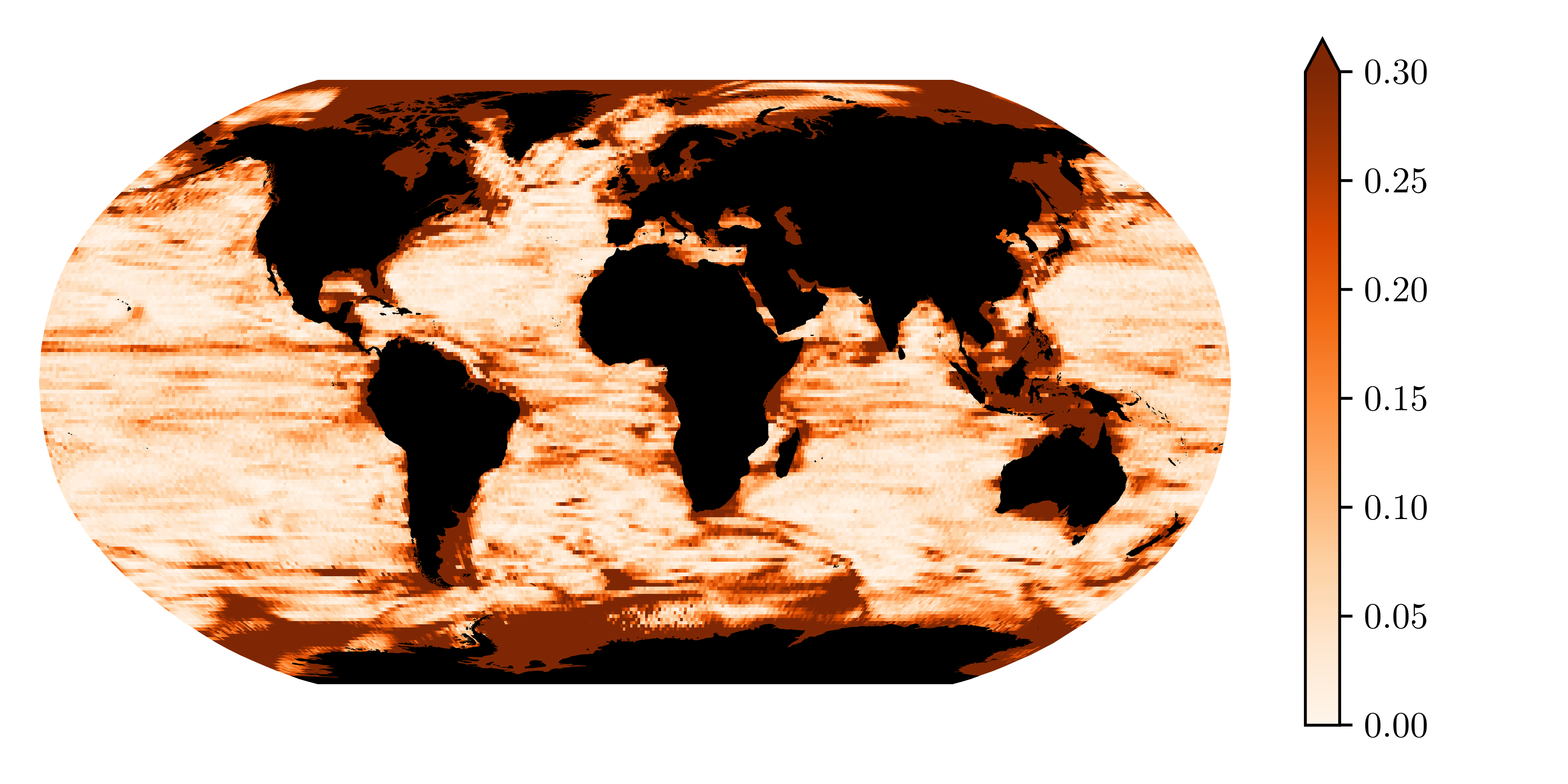}
    \caption{Kullback--Leibler divergence of the single-component MDN model from the full $32$-component MDN model, as a function of initial position. Larger values indicate stronger deviations from Gaussianity in displacements.}
    \label{fig:KL}
\end{figure}

In order to quantify the extent to which the transition density deviates from being Gaussian, and how this varies from one region of the ocean to another, we
computed the Kullback--Leibler (KL) divergence\footnote{The KL divergence of $p$ from $q$, also known as the relative entropy, defined $\infdiv{q}{p}=\int q(x)\log \frac{q(x)}{p(x)} \, \mathrm{d}x$, is a measure of the divergence of a probability density $p$ from a reference probability density $q$ --- often interpreted as the amount of information lost when $p$ is used to approximate $q$.} of the single-component MDN model, which is Gaussian, from the full $32$-component model as a function of initial position. The result is shown in figure~\ref{fig:KL}. Note that, since a closed-form expression for the KL divergence between two Gaussian mixtures is not known~\citep{Cui}, we provide simple Monte Carlo estimates based on $5000$ samples at each of the vertices of a $1^{\circ}\times1^{\circ}$ grid. Where the KL divergence is zero, the two models agree exactly, indicating that displacements are Gaussian. The larger the KL divergence is, the greater the disagreement between the models, and the further from Gaussian the full model is. As a point of reference for interpreting the magnitude of the KL divergence, note that the if $Z_0\sim\mathcal{N}(m_0,\, 1)$ and $Z_1\sim\mathcal{N}(m_1,\, 1)$, then, writing their pdfs as $p_0$ and $p_1$, $\infdiv{p_1}{p_0} = (m_1 - m_0)^2$.
Non-Gaussianity of displacements is likely due primarily to inhomogeneity of ocean velocities --- drifters can explore a range of flow statistics as they move, and the convolved effects of these are reflected in observed displacements.  An alternative explanation is that the underlying velocity field is non-Gaussian --- evidence of non-Gaussian velocities in the North Atlantic has been presented by~\citet{Bracco2000b} and~\citet{LaCasce2005} on the basis of observations from both subsurface current meters and subsurface floats.

As can be seen in figure~\ref{fig:gulf_stream}, the model assigns nonzero probability to drifter displacements intersecting land. This is unavoidable given that the support of the assumed parametric form, that of a Gaussian mixture, extends to infinity; moreover, this may not be entirely spurious, given that some drifters do run aground. In 2012 \citet{Lumpkin2012} reevaluated drifter data to study the causes of drifter deaths. They concluded that approximately $27\%$ of drifter deaths were due to running aground, with a further $10\%$ being picked up by humans, and the remainder failing due to internal faults. Outside of coastal regions this issue is unlikely to have a strong effect on the estimates of displacement statistics considered in section~\ref{displacement_stats}. The implications for drifter simulations are discussed further in section~\ref{drifter_sims}.

\subsubsection{Displacement statistics}\label{displacement_stats}
In this section we present maps of single-particle statistics derived from the model. As a first example, we show the mean of displacements over the $4$-day time increment of our model. We further provide global estimates of lateral diffusivity.

\begin{figure}[htp]
    \centering
    \subfloat[Mean of 4-day zonal displacement (km).]{
        \includegraphics[clip, width=6in]{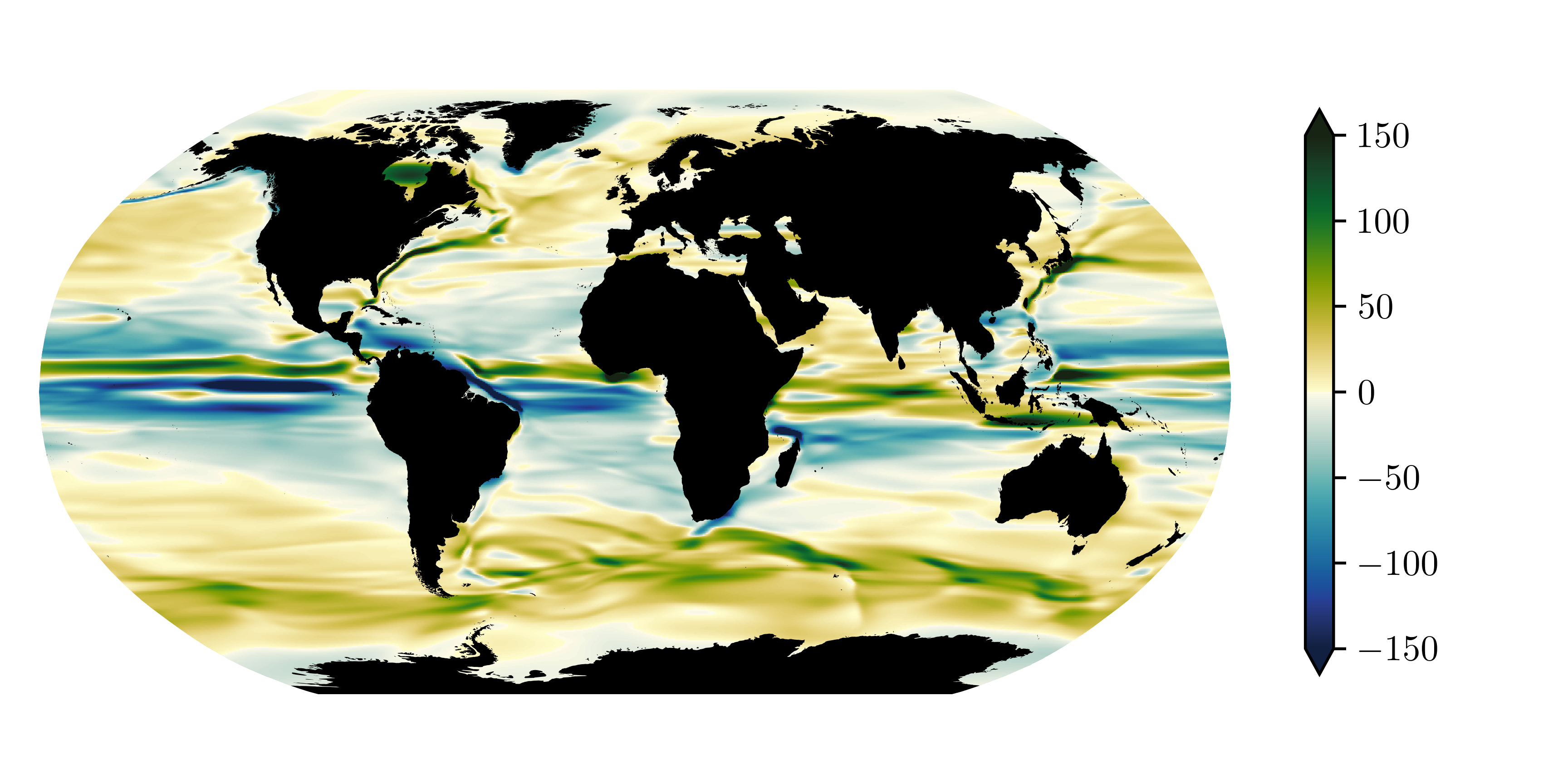}}
    \hspace{0pt}
    \subfloat[Mean of 4-day meridional displacement (km).]{
        \includegraphics[clip, width=6in]{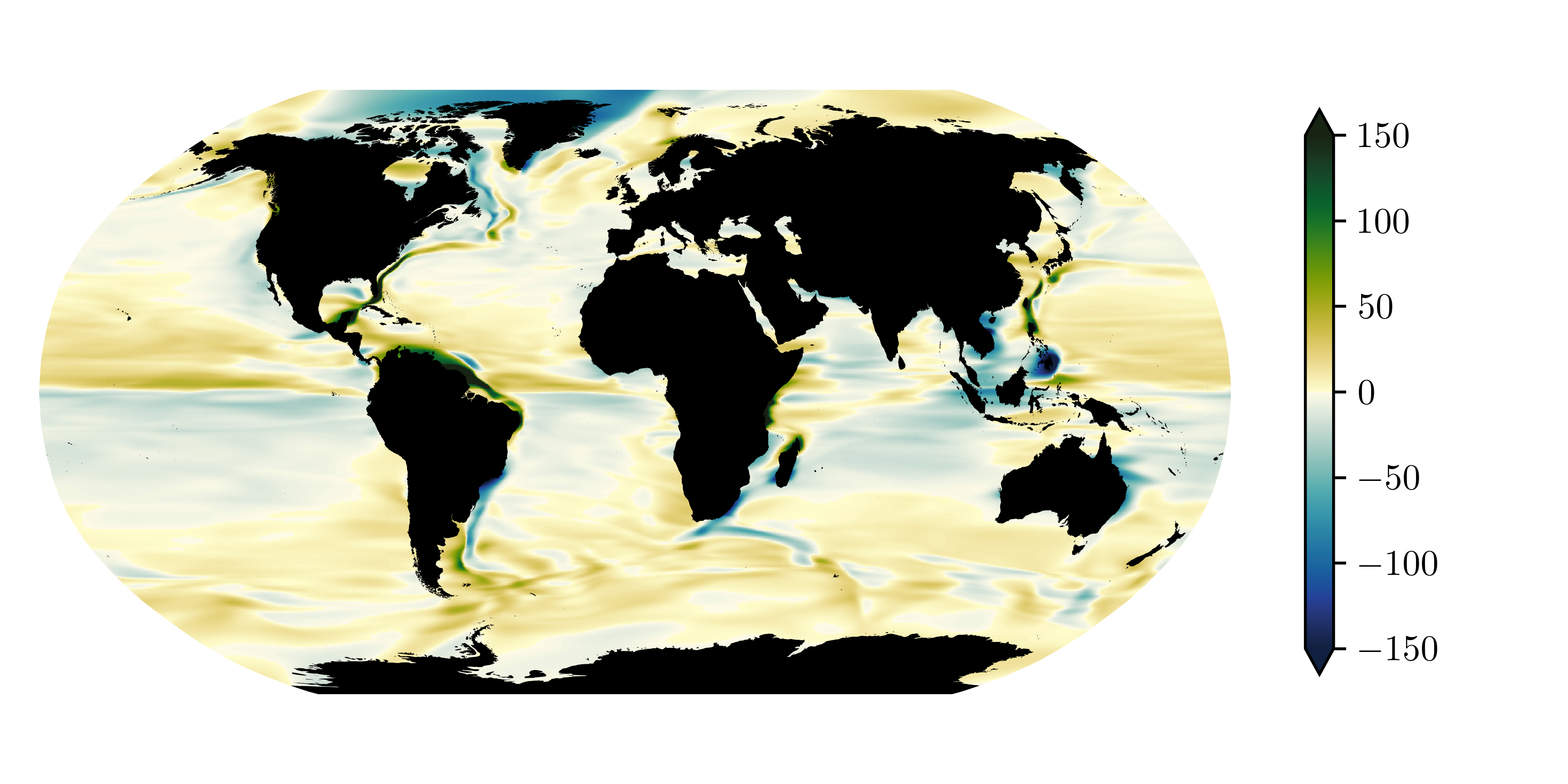}}
    \caption{Mean of displacements from the MDN model, with $\tau =4\text{ days}$, as a function of initial position.}
    \label{fig:mean}
\end{figure}

Figure~\ref{fig:mean} shows the mean of drifter displacements as a function of initial position. While the output of the model is in longitude--latitude coordinates $(\lambda, \phi)$, we apply a simple conversion to kilometres based on a local tangent-plane approximation
\begin{subequations}\label{tangent}
\begin{align}
    \Delta X &= R\, \Delta \phi\\
    \Delta Y &= R\, \Delta \lambda \, \cos{\phi_0},
\end{align}
\end{subequations}
where $R$ is the radius of the Earth at the equator. The imprint of several features of the surface dynamics, such as the western boundary currents and equatorial (counter) currents, is clear.

\begin{figure}[htp]
    \centering
    \subfloat[Scalar estimate of lateral diffusivity, $K^{(4)}$ ($\text{m}^2\text{s}^{-1}$), derived from the MDN model with {$\tau=4$~days}.]{
        \includegraphics[clip, width=6in]{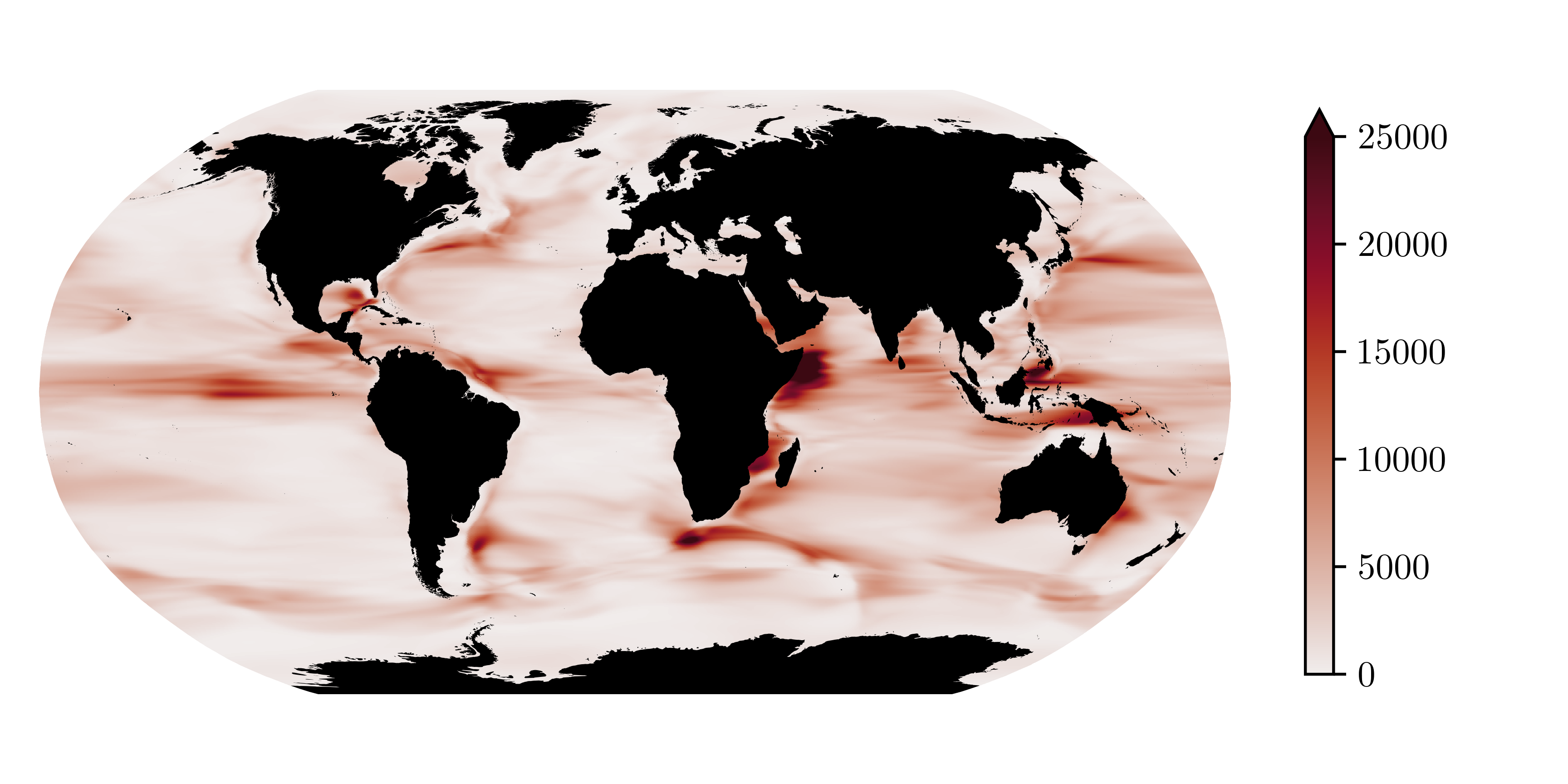}}
    \hspace{0pt}
    \subfloat[Difference of scalar estimates of lateral diffusivity, $K^{(14)} - K^{(4)}$, derived from MDN models with {$\tau=14$~days} and {$\tau=4$~days}, respectively.]{
        \includegraphics[clip, width=6in]{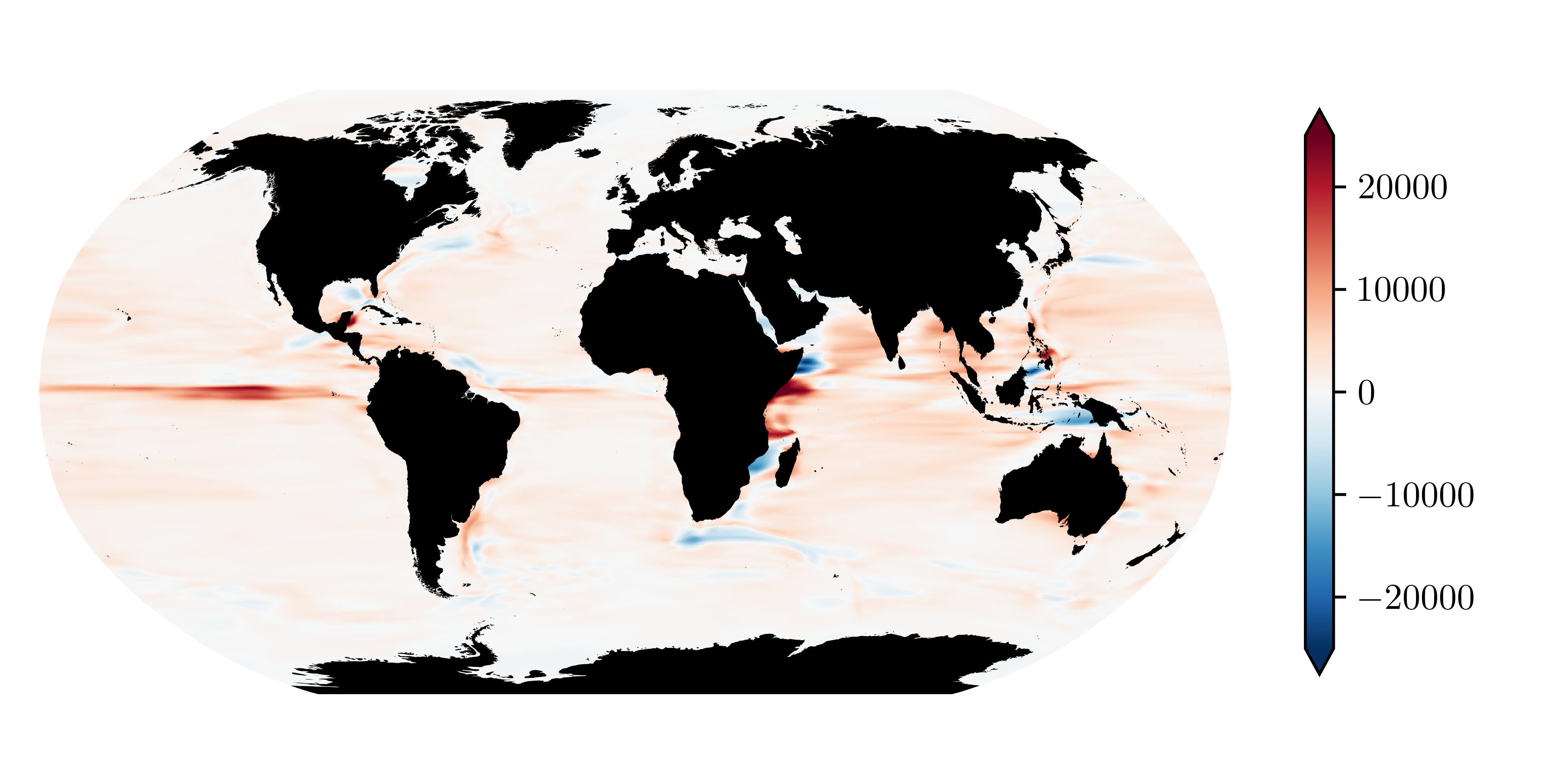}}
    \caption{Global estimate of lateral diffusivity derived from the MDN model of the transition density.}
    \label{fig:diff}
\end{figure}

For the sake of comparison with previous work, we consider the estimation of lateral diffusivity from our model, though we emphasise that by modelling the full transition density, we provide a more accurate description of Lagrangian statistics than can be captured by the familiar advective-diffusive model of dispersion put forward by~\citet{Davis1987, Davis1991}. The estimation of ocean diffusivity by various methods has been the subject of numerous papers~\citep{Oh2000, Zhurbas2003, Zhurbas2004, Klocker2012, Abernathey2013, Klocker2014, Ying}. The estimation of diffusivity from drifter displacements is straightforward only when there exists a suitable sampling time, which is larger than the time for drifter velocities to decorrelate, i.e. for drifter motion to become diffusive, and such that the scale of drifter displacements over that time scale is small relative to spatial variations in the diffusivity. In this case a simple estimate of the lateral diffusivity tensor $\bm{\mathsf{K}}(\bm{x})$ is
\begin{align}\label{eq: diff}
    \bm{\mathsf{K}}(\bm{x}) = \frac{1}{2\,\tau}\,\text{Cov}(\Delta \bm{X}\mid \bm{X}_0=\bm{x}),
\end{align}
where $\tau$ is the suitably chosen time scale, and the conditional covariance is estimated by either one of the approaches sketched in figure~\ref{fig:cond_stats}. Unfortunately, such a time scale may not exist in the ocean, and, if it does exist, it likely varies in space, making its determination difficult. This challenge has been borne out in previous studies~\citep{LaCasce2014, Zhurbas2014}. \citet{Oh2000} proposed a method to circumvent the issues created by inhomogeneity. They proposed to isolate the cross-flow component of the displacement covariance, identified by the minor principal component (the smaller eigenvalue of displacement covariance), and use this to provide a scalar estimate of diffusivity, since the cross-flow component is less affected by shear in the mean flow. In figure~\ref{fig:diff}(a) we provide a similar estimate, derived from the MDN model with $\tau=4$~days,
\begin{align}\label{eq: diff_scalar}
    K(\bm{x}) = \frac{1}{2\,\tau}\, \lambda_2(\bm{x}),
\end{align}
where $\lambda_2(\bm{x})$ is the smallest eigenvalue of
\begin{align}
    \text{Cov}(\Delta \bm{X}\mid \bm{X}_0=\bm{x}) = \sum_i \alpha_i \bm{\mathsf{C}}_i + \sum_i \left[\alpha_i\left(\bm{\mu}_i - \sum_i \alpha_i\bm{\mu}_i\right)\left(\bm{\mu}_i - \sum_i \alpha_i\bm{\mu}_i\right)^T\right].
\end{align}
The result agrees very well with estimates provided by~\citet{Zhurbas2004} for the Atlantic and Pacific oceans. Figure~\ref{fig:diff}(b) shows the difference of estimates of the form~\eqref{eq: diff_scalar} with $\tau=14$~days and $\tau=4$~days, respectively. In many areas, the diffusivity estimates are slightly amplified by taking a larger time lag $\tau$, with greater differences visible in some particularly energetic regions; however, the effect is indeed much weaker than that observed with analogous along-flow diffusivity estimates derived from the largest eigenvalue of the displacement covariance matrix.

Leaving aside the challenges of estimating diffusivity from displacements, which are common to all methods, we highlight as this point some advantages to our approach. Using the MDN model, trained with maximum likelihood and effectively regularised by the use of early-stopping, removes the difficulty of tuning the resolution of bins. Instead, the effective resolution of our mean displacement and diffusivity estimates is set automatically by the resolution of the data, and is free to vary optimally in space. This allows us to produce at once global estimates, which resolve well-sampled flow features very well and are forgiving in regions where data is relatively sparse, with the exception of very high-latitude regions, where there is simply no data to constrain the model.

\subsubsection{Drifter simulations}\label{drifter_sims}

In this section we demonstrate the simulation of drifter trajectories using the MDN model as the basis for a discrete time Markov process model. In a discrete-time setting, assuming Markovianity means assuming that $p(\bm{X}_{n+1}\mid \bm{X}_n,\, \bm{X}_{n-1},\, \cdots)=p(\bm{X}_{n+1}\mid \bm{X}_n)$. In this case, sampling trajectories amounts to repeatedly sampling displacements in sequence according to the transition density, since, given the current position, displacements are statistically independent of previous positions.

A complication of simulating drifters in this way is that, for reasons discussed above, drifters can hit land. In this work we do not attempt to model the beaching of drifters, since it is not clear that the Global Drifter Program dataset contains sufficient reliable information --- in particular, it remains a challenge to determine whether drifters have run aground or not~\citep{Lumpkin2012}. To exclude the possibility of running aground in our drifter simulations we implement a simple rejection sampling scheme, wherein displacements sampled from the transition density which would bring a drifter on land are rejected, and a new displacement is sampled until a displacement which keeps the drifter in the ocean is drawn. This amounts to sampling according to the conditional density $p(\Delta\bm{X}\mid\bm{X}_0, \bm{X}_0 + \Delta\bm{X} \notin \text{land})$, and is equivalent to the standard practice when using transition matrix models of restricting the domain considered to the ocean and normalising probability estimates correspondingly. To determine whether a proposed new position is on land, we check intersection with a $110\mathrm{m}$-resolution land mask.

We simulated the evolution of a set of drifters initialised on the vertices of a $2^{\circ}\times2^{\circ}$ grid for a period of $10$ years. Note that the evolution of each drifter is simulated independently. This means that multi-particle statistics that would characterise the joint evolution of drifters released simultaneously in the ocean are not represented and, in particular, that the current model is not appropriate for simulating the release of a cloud of tracer particles on short time scales; however, it can be expected to represent the behaviour of drifters or buoyant tracers over large spatial and temporal scales. Similar experiments, carried out by~\citet{Maximenko} and~\citet{vanSebille2012} using transition matrix models trained on Global Drifter Program data, studied the clustering of simulated drifters due to near-surface convergence and the formation of so-called garbage patches, including the North Pacific Garbage Patch~\citep{Moore} and others corresponding to the other subtropical ocean gyres. The simulations of~\citet{vanSebille2012} showed a further cluster in the Barents Sea which formed only after several decades.

\begin{figure}
    \centering
    \includegraphics[width=6in]{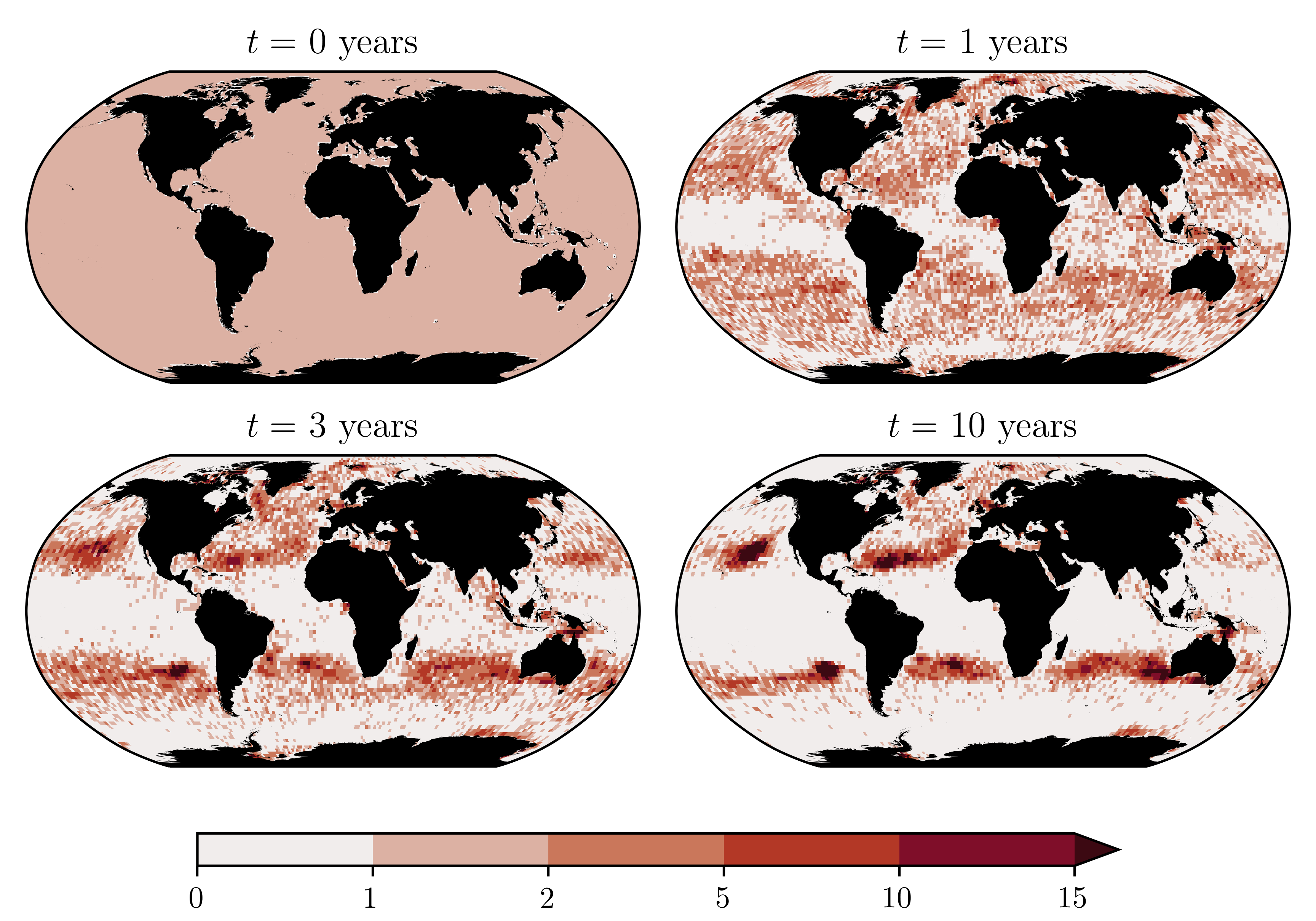}
    \caption{Histograms of simulated drifters initially and after one, three, and ten years of evolution under the MDN model, respectively.}
    \label{fig:hists}
\end{figure}

The results of our model simulation are largely in agreement with these previous studies. The distribution of the simulated drifters is shown in figure~\ref{fig:hists} at the beginning of the simulation and after one, three, and ten years of evolution under the MDN model. After one year the drifters have become relatively sparse in equatorial regions. After three years clusters in the subtropical gyres have begun to appear, and after ten years, these are very well defined. Smaller clusters are also seen to appear, notably in the North Sea and in the seas south of Papua, as well as in some high latitude regions including along the west coast of Greenland and off Antarctica around $100$ -- $130^{\circ}$ E.. Validating these clusters, that is, assessing whether marine debris is likely to accumulate in these areas, is difficult, because in situ observations remain sparse~\citep{Ryan2009}. It may be that the dynamics in these regions, which are poorly sampled by GDP drifters, are simply underresolved by the MDN model, leading to spurious convergence zones. We note, for example, in figure~\ref{fig:mean} that mean displacements do not appear to represent the detail of known currents in the southern portion of the North Sea, which is not visited by drifters in the GDP data (see figure~\ref{fig:drifters}). In general, as is true for any data-driven model, caution should be exercised when interpreting model outputs in regions where data is lacking.

\section{Conclusions}\label{sec: discussion}

This work demonstrates the use of conditional density estimation, and, in particular, stochastic neural networks, in a fluid dynamical problem, namely that of diagnosing single-particle statistics from trajectory data. We show how such probabilistic models are useful both as emulators, and as an indirect means of estimating conditional statistics. By operating in the framework of probabilistic modelling we are able to appeal to the extensive literature on statistical inference, probabilistic forecasting, model comparison and validation, and thereby avoid ad hoc choices of loss functions and performance metrics. Our model is compared, using a probabilistic scoring rule, to alternative models, including a Markov chain model used extensively in the literature, and is shown to outperform these, both globally and in three specific regions.

By modelling the single-particle transition density of surface drifters, we gain estimates of a range of conditional statistics simultaneously, which capture the occurrence of strongly non-Gaussian statistics in some areas of the ocean. We provide global maps of mean displacement and lateral diffusivity, but emphasise that these examples provide only a limited summary of the information contained in the transition density; further statistics, including higher moments of displacements can readily be computed from our model. Interpreted as the basis for a discrete-time Markov process, our model is also used to simulate the evolution of a set of drifters seeded globally on a uniform grid, and shows the emergence of clusters of drifters in the subtropical gyres, in agreement with previous work on the formation of garbage patches.

The approach espoused in this work is equally applicable to other problems in fluid dynamics and oceanography. One example is the estimation of structure functions from either Eulerian or Lagrangian velocity data. Another is the estimation of multi-particle statistics, such as relative dispersion, via modelling of multi-particle transition densities. Yet another is the learning of stochastic parameterisations in climate/atmosphere/ocean models. \citet{Guillaumin} made progress on the parameterisation of subgrid momentum forcing in an ocean model with a single-component MDN model, but the approach is applicable more broadly, e.g. to the parameterisation of subgrid transport.

In this work we have largely neglected the need to quantify uncertainty in model parameters and to incorporate prior knowledge in our modelling. These needs would be met by a Bayesian approach, where, instead of estimating parameters by maximum likelihood, we apply Bayesian inference to obtain posterior distributions on parameters, which account for prior knowledge. Indeed, all of the results presented herein would benefit from uncertainty quantification. In the case of conditional statistics, a Bayesian approach would, e.g., allow to identify where there is not enough data to inform  reliable estimates of lateral diffusivity; and in general, incorporating prior knowledge may help to regularise our model of the transition density, so that, in the case of drifter simulations, spurious convergence zones can be avoided. The application of Bayesian inference to MDNs remains challenging, but we consider this an important future direction.



\small
\vspace{0.5em} \noindent \textbf{Acknowledgements.} I am grateful to Jacques Vanneste, James Maddison and Aretha Teckentrup for their advice, input and overall support of this work. I also thank Dhruv Balwada for helpful discussions. Thanks are also due to the reviewers for their suggestions which have improved the manuscript.

\vspace{0.5em} \noindent \textbf{Funding.} The author was supported by the MAC-MIGS Centre for Doctoral Training under EPSRC grant EP/S023291/1.
This work used the Cirrus UK National Tier-2 HPC Service at EPCC \\(\href{http://www.cirrus.ac.uk}{www.cirrus.ac.uk}) funded by the University of Edinburgh and EPSRC (EP/P020267/1).

\vspace{0.5em} \noindent \textbf{Declaration of interests.} The author reports no conflict of interest.

\vspace{0.5em} \noindent \textbf{Data availability statement.} The code required to reproduce the results herein is available at\\
\href{https://doi.org/10.5281/zenodo.7737161}{doi.org/10.5281/zenodo.7737161}, along with the trained MDN model and a Jupyter Notebook which demonstrates its use. The processed GDP data used and drifter simulation data are available at \\\href{https://doi.org/10.7488/ds/3821}{doi.org/10.7488/ds/3821}.
\normalsize

\bibliography{main}{}

\end{document}